\documentclass[aps,prb,twocolumn,superscriptaddress,citeautoscript,showpacs,floatfix]{revtex4}

\usepackage{graphicx}
\usepackage{bm}
\usepackage{times}

\newcommand{\smeq}{\! = \!}

\newcommand{\smpl}{\! + \!}
\newcommand{\smmi}{\! - \!}

\newcommand{\kt}{k_{\text{B}}T}

\newcommand{\be}{\begin{equation}}
\newcommand{\ee}{\end{equation}}
\newcommand{\bea}{\begin{eqnarray}}
\newcommand{\eea}{\end{eqnarray}}

\begin{document} 



\title{Level Spacings in Random Matrix Theory and Coulomb Blockade Peaks in Quantum Dots}

\author{Damir Herman}
\altaffiliation{Myeloma Institute for Research and Therapy,
University of Arkansas for Medical Sciences, Little Rock AR 72205.}
\affiliation{Department of Physics, Case Western Reserve University,
  Cleveland,  Ohio 44106-7079} 
\author{T. Tzen Ong}
\altaffiliation{Dept. of Applied Physics, Stanford University, Stanford CA 94305.}
\affiliation{Department of Physics, Duke University, Box 90305, Durham,
  North Carolina 27708} 
\author{Gonzalo Usaj}
\affiliation{Department of Physics, Duke University, Box 90305, Durham,
  North Carolina 27708}
\affiliation{Instituto Balseiro and Centro At\'{o}mico Bariloche, Comisi\'{o}n 
Nacional de Energia At\'{o}mica, (8400) Bariloche, Argentina.}
\author{Harsh Mathur}
\affiliation{Department of Physics, Case Western Reserve University,
  Cleveland,  Ohio 44106-7079} 
\author{Harold U. Baranger}
\affiliation{Department of Physics, Duke University, Box 90305, Durham,
  North Carolina 27708} 

\date{August 17, 2007} 

\begin{abstract}
We obtain analytic formulae for the spacing between conductance peaks in the Coulomb blockade regime, based on the universal Hamiltonian model of quantum dots. New random matrix theory results are developed in order to treat correlations between two and three consecutive spacings in the energy level spectrum. These are generalizations of the Wigner surmise for the probability distribution of single level spacing. The analytic formulae are shown to be in good agreement with numerical evaluation.
\end{abstract} 

\pacs{73.23.Hk,05.45.Mt,73.21.La,02.10.Yn}

\maketitle

\section{Introduction}

The Coulomb blockade of electrons has been a remarkable tool for probing fundamental physics in nanoscale systems \cite{review,IhnBook,glazmanreview,oregreview}.  In a semiconductor quantum dot (QD), weakly coupled to leads via tunneling, the effect is manifested as a series of spikes in the conductance of the device as a function of a gate voltage which controls the number of electrons on the QD.  In the quantum Coulomb blockade regime (defined below), the spacing between the spikes is determined by the ground state energy of the QD. Experimentally, the spacing shows mesoscopic fluctuations, and there have been several detailed studies of the statistical distribution of the peak spacings \cite{sivan,patel1,patel2,simmel,luscher01,ihn02,fuhrer03,ong}.

The distribution of peak spacings results from the interplay of electron-electron interaction and randomness. A typical semiconductor QD in the experiments of Ref.\,\onlinecite{patel1} consists of a droplet of several hundred electrons confined to a two dimensional region of a few tenths of a micron in size.  In the absence of electron-electron interaction, the motion of the electrons would be randomized by impurity scattering or by chaotic scattering from the boundaries of the QD; hence, it is expected \cite{BohigasLesHouches,EfetovBook} that the single particle energy levels in the absence of interaction would be described by random matrix theory (RMT) \cite{mehta}.  In a real QD, however, it is necessary to take the additional effects of interactions into account. This is indicated by the failure of (essentially) non-interacting models to account for the observed distribution of Coulomb blockade peak spacings \cite{sivan,patel1}.

The interplay of electron-electron interaction and randomness is a notoriously hard problem. Nonetheless, due to the finite size of the QD, it has been possible to make significant progress. In particular, it is now believed that the energy levels of a weakly interacting QD are described statistically by a ``universal Hamiltonian'' (see Refs.\,\onlinecite{glazmanreview} and\ \onlinecite{oregreview}, and references therein).  According to this model, each QD is characterized by a set of single particle orbitals with single particle energies and wavefunctions distributed according to the appropriate ensemble of RMT. Given the occupation numbers of these single particle orbitals, the non-interacting contribution to the energy of a QD follows directly.  The interaction contribution is determined entirely by the net charge on the QD (the ``charging energy'') and its total spin (the ``exchange energy''). Thus the universal Hamiltonian is
\begin{equation}
H \smeq  \sum_{i \sigma} \epsilon_i \hat{n}_{i \sigma}
\smpl  \frac{ e^2 }{ 2 C } \delta \hat{N}^2 
\smmi J \hat{ {\mathbf S} }^2 \;,
\label{eq:universal}
\end{equation}
where $ \hat{n}_{i \sigma} \smeq  $ 0 or 1 is the occupation number of orbital $i$ with spin $ \sigma \smeq\,  \uparrow {\rm or} \downarrow$, $ \delta \hat{N} $ is the number of excess electrons on the QD, and $ \hat{ {\mathbf S} } $ is the total spin-operator. $C$ denotes the capacitance of the QD, and $J$ is the exchange constant.  Strictly, there is a fourth term in the universal Hamiltonian corresponding to interaction in the Cooper channel. However, this term vanishes when time reversal symmetry is broken by the application of a magnetic field, and, even when present, it is significantly smaller than the others and may be neglected \cite{glazmanreview}.

It should be emphasized that the universal Hamiltonian does not help calculate the energy levels of any particular QD. Rather it provides a description of the universal statistical features of the levels of all QDs that belong to the universality class under consideration.

The form of the universal Hamiltonian is dictated by the few symmetries that remain to a random system such as a QD \cite{kurland}. For example, Eq.~(\ref{eq:universal}) applies provided the QD has spin-rotation invariance (no spin-orbit or spin-flip scattering). If in addition time reversal symmetry is intact, the single particle levels are distributed according to the orthogonal ensemble of RMT; if time reversal symmetry is broken (by a magnetic field, for example), the unitary ensemble applies.

If the exchange constant is set to zero, the universal Hamiltonian reduces to the old constant interaction model which has proven unsuccessful in accounting for the observed peak spacing distribution.  It is instructive to compare the physics of the two models.  According to the constant interaction model, each filled QD level is doubly occupied in the ground state, except for the top level, which would be singly occupied if the total number of electrons in the dot were odd. Thus the total spin of the ground state is zero or one-half depending on whether the number of electrons is even or odd. Within the universal Hamiltonian model, higher spin ground states are possible: the higher single particle energy of these states is offset by the exchange energy which favors parallel alignment of spins \cite{brouwer,baranger}. Indeed if the exchange constant is above a certain threshold, the QD should have a ferromagnetic ground state that is fully spin-polarized \cite{kurland,AndKam98}.  Here we restrict our attention to the paramagnetic regime in which the ground state may have a substantial spin but is not fully polarized.

It is also instructive to compare the physics of QDs to that of atoms. Atoms can be approximately understood by considering that the electrons occupy a set of self-consistent hydrogen-like orbitals. The precise filling of the incomplete shells (which determines the total spin of the atomic ground state) is then controlled by the interplay of electron-electron interaction and spin-orbit coupling encoded in Hund's rules. The universal Hamiltonian provides a comparable description of the energy levels of weakly interacting QDs. The key difference between atoms and QDs is that for atoms the confining potential is spherically symmetric whereas for QDs it is essentially random.

The universal Hamiltonian is found to give a good quantitative account of the experiments of Ref.\,\onlinecite{patel1} when finite temperature effects and small non-universal corrections are included \cite{ong,ullmo01,gonzalo,UsajB02,AlhassidM02}. Here, we  obtain analytic formulae for the distribution of conductance peak spacings in the low temperature and large QD limit. Though the distribution has been evaluated numerically before \cite{ullmo01,gonzalo,UsajB02,AlhassidM02}, analytic expressions should prove helpful in the analysis of experimental data.

The quest to develop such analytic expressions raises some interesting problems in RMT.  Within the constant interaction model, the distribution of peak spacings, $\Pi(s)$, is controlled by the distribution of the spacing between consecutive energy levels, denoted $P_1 (\Delta)$. Level spacing distributions are notoriously hard to calculate but by now this distribution is well understood \cite{mehta}.  The exact analytic expression for $P_1 (\Delta)$ is far too complicated to be useful in practice. Fortunately, the celebrated Wigner surmise provides a simple and remarkably accurate approximation to the exact result. Within the universal Hamiltonian model, however, the peak spacing distribution is controlled by the joint probability distribution of the spacings between several consecutive levels, which we shall denote as $P_2(\Delta_1, \Delta_2)$, $P_3 (\Delta_1, \Delta_2, \Delta_3)$, etc. As for the single spacing distribution, the exact expressions for the joint spacing distributions are too cumbersome for practical use. 

We have, therefore, developed approximate expressions analogous to Wigner's surmise for the joint level spacing distributions and used these to obtain formulae for the conductance peak spacing distribution. In addition, we have developed new numerical techniques for evaluating the exact expressions which are much more efficient than previously known methods. The approximate analytic expressions are in excellent agreement with the numerics.

To our knowledge there has been no previous investigation of these joint spacing distributions in a physics context \cite{Orsay99}. Number theorists have studied a quantity dubbed the nearest neighbor spacing distribution in connection with the zeros of Riemann's zeta function \cite{ForrOdlyzko}.  That distribution is closely related to $P_2 (\Delta_1, \Delta_2)$, and, indeed, the zeta function work, as well as other known results, provide a useful test of the accuracy of the approximate formulae we develop.

The rest of the paper is organized as follows. First, we start with the main results of interest from a QD point of view: Section II gives our analytic results for the peak spacing distribution. Then, in Section III we discuss the RMT quantities on which the peak spacing distribution is based. The approximate expressions for the joint distribution of several consecutive levels are developed here. Section IV presents our exact results for these RMT quantities and compares them with the previously given surmises. Finally, in Section V we summarize and conclude.

\section{Peak Spacing Distribution \label{sec:peakspacing}}

The peaks in the conductance occur at values of the gate voltage for which the number of electrons on the dot changes by one. In the quantum Coulomb blockade regime, the spacing between the $N\smmi1 \!\rightarrow\! N$ and $N \!\rightarrow\! N \smpl  1$ peaks is
\begin{equation}
\delta_2 \equiv E_{N+1} \smpl  E_{N-1} \smmi 2 E_{N} \;,
\label{eq:Delta2}
\end{equation}
where $ E_N $ denotes the ground state energy of the QD with $N$ electrons. The sequence of peak spacings is sometimes called the addition spectrum of the QD. For later use it is convenient to define the shifted peak spacing
\begin{equation}
s \smeq  \delta_2 - \frac{ e^2 }{ C }\;.
\label{eq:delta2}
\end{equation}

It is instructive to first consider the constant interaction model.  
For $N$ even, the $N\smmi1 \!\rightarrow\! N$ transition involves adding an electron to level $N/2$, while in the $N \!\rightarrow\! N \smpl  1$ transition an electron is added to level $N/2\smpl1$. Thus, for an odd-even-odd transition (for brevity, an even spacing hereafter)
%
\begin{equation}
s\smeq  \Delta\;,
\label{eq:constinteractioneven}
\end{equation}
where $ \Delta \smeq \epsilon_{N/2+1} \smmi \epsilon_{N/2}$ is the spacing between the two levels.  Similarly, for an even-odd-even transition (an odd spacing hereafter)
\begin{equation}
s \smeq  0 \;.
\label{eq:constinteractionodd}
\end{equation}
Thus in the constant interaction model, the peak spacing distribution is bimodal: the even spacings are distributed in the same way as the single particle level spacings while the odd spacing distribution is a delta function spike at $ s \smeq  0 $.

Let us now consider the case of the universal Hamiltonian model, Eq.~(\ref{eq:universal}).  For even $N$, the state with the lowest $N/2$ single particle levels doubly occupied has total spin $S_N\smeq0$. It is therefore an eigenstate of the universal Hamiltonian although, as we shall see, it is not necessarily the ground state. Next, consider the four states obtained by promoting an electron from the highest occupied level to the lowest unoccupied one (the occupations here are stated relative to the constant interaction ground state). The single-particle energy of these states is greater by $\Delta$, and they can be combined into a spin triplet and singlet. We have thus identified five eigenstates of the universal Hamiltonian: the constant interaction ground state with energy $ E_0 $, the degenerate triplet states with energy $ E_0 \smpl  \Delta \smmi 2 J $ and the singlet state with energy $ E_0 \smpl  \Delta $. Proceeding in this way we can construct all the eigenstates of the universal Hamiltonian from the excitations of the constant interaction model.

For $ J\smeq0 $, the constant interaction ground state is the true ground state, but for $ J\! >\! \Delta/2 $ the triplet states become lower in energy. In principle a state with still higher spin might have even lower energy: such states are costlier in terms of single particle energy but have large (negative) exchange contributions to their energy. One might expect that in the paramagnetic limit $ J\!<\! \langle \Delta \rangle $ (the mean level spacing) the single particle energy would dominate, and a high spin state would be unlikely. Indeed for $ J/ \langle \Delta \rangle $ as large as $0.5 $ we have verified that, in systems without time-reversal symmetry, approximately $ 98\% $ of the ground states for even $N$ have $ S_N\smeq0 $ or $ S_N\smeq1 $. Similarly, $ S_N\smeq\frac{1}{2} $ and $ S_N\smeq\frac{3}{2} $ are the dominant spin states for $ N $ odd.  Hence, in the following we will consider only the competition between the two lowest spin states. 

Now consider the peak spacing for an even spacing.
For simplicity, first assume that $ S_{N-1}\smeq S_{N+1} \smeq\frac{1}{2}$ and $S_N \smeq 0$ or $1$; in this case the spacing is given by
\be
s \smeq \left\{  
\begin{array}{ll}
\Delta \smmi \frac{3}{2} J & {{\rm for }}\, \Delta\! >\! 2 J\\
&\\
\frac{5}{2} J \smmi \Delta & {{\rm for}} \, \Delta\! <\! 2 J\,.
\end{array}\right.
\label{eq:universaleven}
\ee
Note that $ s \!\ge\! J/2 $. The cumulative probability that the spacing is smaller than a certain value
is
\be
F ( s ) \smeq \left\{
\begin{array}{ll}
 0 & {{\rm for}}\,\,\, s\! <\! \frac{1}{2} J \\
&\\
\displaystyle{\int_{\frac{5}{2}J - s}^{\frac{3}{2}J + s}} P_1( \Delta )\,d\Delta & {{\rm for}}\,\,\, \frac{1}{2} J\! <\! s\! <\! \frac{5}{2} J \\
&\\
\displaystyle{\int_{0}^{\frac{3}{2}J + s}} P_1( \Delta )\,d\Delta & {{\rm for}} \,\,\, \frac{5}{2} J\! <\! s \;.
\end{array}\right.
\label{eq:cumuluniversaleven}
\ee
Here $ P_1( \Delta ) $ denotes the single particle level
spacing distribution, for which both the exact result 
and an extremely accurate approximation---Wigner's surmise---are
known in RMT.\cite{mehta}
By differentiation we can convert the cumulative distribution to the
peak spacing probability distribution
\be
\Pi_{{\rm even}} ( s ) \smeq \left\{
\begin{array}{ll}
 0 & {{\rm for}} \,\, s\! <\! \frac{1}{2} J \\
&\\
P_1 \left( \frac{5}{2} J\smmi s\right) \smpl P_1 \left( \frac{3}{2} J \smpl s \right) & {{\rm for}}\,\, \frac{1}{2} J\! <\!s\!<\! \frac{5}{2} J\\
&\\
P_1 \left( s \smpl  \frac{3}{2} J \right)& {{\rm for}}\,\, \frac{5}{2} J\! <\!s\;.
\end{array}\right.
\label{eq:densityeven}
\ee
Note that there is a sharp jump in the distribution $ \Pi_{{\rm even}} $ at the left edge of its support $ ( s \smeq J/2 )$, in contrast to the smooth behavior in the constant interaction model. Also $ \Pi_{{\rm even}} $ is continuous at $ 5J/2 $ but kinked in systems with time-reversal symmetry [see Eq.~(\ref{eq:wsgoe}) below].

The odd spacing is similar but more tedious to analyze because there are
four cases to consider: the QD may have either spin zero or
spin one in both initial and final states. We denote the spacing between
the $ [(N\smmi1)/2]^{{\rm th}} $ level and the level above it $ \Delta_1 $ and
between the $ [(N\smpl1)/2]^{{\rm th}}$ level and the level above that
$ \Delta_2 $. Straightforward analysis then shows that
\be
s \smeq  \left\{
\begin{array}{ll}
 \frac{3}{2} J & {{\rm for}}\,\, \Delta_1\!>\! 2 J \textrm{~and~} \Delta_2\! >\! 2 J\\
&\\
\Delta_2 \smmi \frac{1}{2} J & {{\rm for}}\,\, \Delta_1\! >\! 2 J \textrm{~and~} \Delta_2\! <\! 2 J\\
&\\
\Delta_1\smmi \frac{1}{2} J & {{\rm for}}\,\, \Delta_1\! <\! 2 J \textrm{~and~} \Delta_2\! >\! 2 J\\
& \\
\Delta_1 \smpl  \Delta_2 \smmi \frac{5}{2} J & {{\rm for}}\,\, \Delta_1\! <\! 2 J \textrm{~and~} \Delta_2\! < \!2 J \;.
\end{array}\right.
\label{eq:universalodd}
\ee
Note $ -\frac{5}{2}J\!\leq\!s\!\leq\!\frac{3}{2}J $. Provided $ \Delta_1\! >\!2 J $ and $ \Delta_2\! >\!2 J $ (always true in
the constant interaction limit of $ J\! \rightarrow\! 0 $) all odd
peak spacings have the same shifted value $ s \smeq \frac{3}{2}J$
independent of the values of the level spacings.
However if these conditions are not met the spacing does depend
on the values of $ \Delta_1 $ and $ \Delta_2 $. A short 
calculation reveals
\be
\Pi_{{\rm odd}} (s ) \smeq  \delta\! \big( s\smmi \frac{3}{2} J \big) \!\!
\int_{2 J}^{\infty}\!\!\int_{2 J}^{\infty}\! P_2( \Delta_1, \Delta_2 ) d \Delta_1 d \Delta_2 
\smpl C ( s )\,.
\label{eq:densityoddi}
\ee
Here $ P_2 ( \Delta_1, \Delta_2 ) $ is the joint probability distribution
of two consecutive level spacings, for which we will derive an accurate
approximation analogous to the Wigner surmise in Section \ref{levelspacing}. 
The continuum distribution is given by 
\be
C ( s ) \smeq \left\{
\begin{array}{ll}
0 & {\rm for}\,\, s\! <\! - \frac{5}{2} J\,\, {\rm or} \,\, s\! >\! \frac{3}{2} J\\
&\\
\displaystyle{\int_0^{s +\frac{5}{2}J}}P_2 \left(\mu, \tilde{\mu} \right)\, dv
& {\rm for}\,\, - \frac{5}{2} J\! <\! s\! <\! - \frac{1}{2} J\\
&\\
\displaystyle{\int_0^{\frac{3}{2}J - s}}\!  P_2 \left( \mu,\tilde{\mu} \right) dv+&\!\!2 \displaystyle{\int_{2J}^{\infty}} \! P_2 \left(v, s \smpl  \frac{J}{2}  \right)\,dv\\
&\\
& {\rm for} \,\,- \frac{1}{2} J\! <\! s\! <\! \frac{3}{2} J
\end{array}\right.
\label{eq:densityoddii}
\ee
with $ \mu\smeq \frac{5}{4}J\smpl \frac{1}{2}(s\smpl v)$ and $ \tilde{\mu}\smeq \frac{5}{4}J\smpl\frac{1}{2}(s\smmi v)$.  

Eqs.~(\ref{eq:densityeven}), (\ref{eq:densityoddi}), and (\ref{eq:densityoddii}) are the expressions for the Coulomb blockade peak spacing distributions when it is assumed that the QD cannot have a ground state spin greater than one. These expressions apply whether time-reversal symmetry is intact or broken: the only difference in the two cases is that it is necessary to use the level spacing distributions $ P_1 ( \Delta ) $, $ P_2( \Delta_1, \Delta_2 ) $ for the Gaussian orthogonal ensemble (GOE) when time-reversal symmetry is intact and for the Gaussian unitary ensemble (GUE) when it is broken.

\begin{table}[t]
\begin{ruledtabular}
\begin{tabular}{|c|c|c|} \hline
 $(S_{N-1},S_N,S_{N+1})$ & Spacing & Condition \\ \hline
$(\frac{1}{2},0,\frac{1}{2})$ & $s \smeq  \Delta_2 \smmi \frac{3}{2}J$ & $\Delta_1 \smpl  \Delta_2 \!\geq\! 3J$ \\ 
&&$\Delta_2 \!\geq\! 2J$ \\
&$\frac{1}{2} J\!\leq\!s$&$\Delta_2 \smpl  \Delta_3 \!\geq\! 3J$ \\ \hline
$(\frac{1}{2},1,\frac{1}{2})$ & $s \smeq  -\Delta_2 \smpl  \frac{5}{2}J$ & $\Delta_1 \smpl  \Delta_2 \!\geq\! 3J$ \\
 & & $\Delta_2 \!\leq\! 2J$ \\
 &$ \frac{1}{2}J\!\leq\!s\!\leq\! \frac{5}{2}J$ & $\Delta_2 \smpl  \Delta_3 \!\geq\! 3J$\\ \hline
$(\frac{1}{2},1,\frac{3}{2}) $ & $s \smeq  \Delta_3\smmi \frac{1}{2}J$ & $\Delta_1 \smpl  \Delta_2 \!\geq\! 3J$ \\
 & & $\Delta_2 \!\leq\! 2J$ \\
 &$ -\frac{1}{2}J\!\leq\!s\!\leq\! \frac{5}{2}J$ & $\Delta_2 \smpl  \Delta_3 \!\leq\! 3J$ \\ \hline
$(\frac{3}{2},1,\frac{1}{2})$& $s \smeq  \Delta_1 \smmi \frac{1}{2}J$ & $\Delta_1 \smpl  \Delta_2 \!\leq\! 3J$ \\
 & & $\Delta_2 \!\leq\! 2J$ \\
 &$ -\frac{1}{2}J\!\leq\!s\!\leq\! \frac{5}{2}J$ & $\Delta_2 \smpl  \Delta_3 \!\geq\! 3J$ \\ \hline
$(\frac{3}{2},1,\frac{3}{2})$ & $s \smeq  \Delta_1 \smpl  \Delta_2 \smpl  \Delta_3 \smmi \frac{5}{2}J$
&$\Delta_1 \smpl  \Delta_2 \!\leq\! 3J$ \\
&&  $\Delta_2 \!\leq\! 2J$ \\
 & $ -\frac{5}{2}J\!\leq\!s\!\leq\! \frac{7}{2}J$& $ \Delta_2 \smpl  \Delta_3 \!\leq\! 3J$ \\ \hline
&&\\ \hline
 $(0,\frac{1}{2},0)$ & $s \smeq \frac{3}{2}J$ & $\Delta_1 \!\geq\! 2J$ \\
 & & $\Delta_2 \!\geq\! 2J $ \\
 && $\Delta_1 \smpl  \Delta_2 \!\geq\! 3J$ \\ \hline
$(0,\frac{1}{2},1)$ & $s \smeq  \Delta_2 \smmi \frac{1}{2}J$ & $\Delta_1 \!\geq\! 2J$ \\
 & & $\Delta_2 \!\leq\! 2J $ \\
 & $-\frac{1}{2}J \!\leq\!s\!\leq\! \frac{3}{2}J$& $\Delta_1 \smpl  \Delta_2 \!\geq\! 3J$ \\ \hline
 $(1, \frac{1}{2}, 0)$ & $s \smeq  \Delta_1 \smmi \frac{1}{2}J$ & $\Delta_1 \!\leq\! 2J$ \\
& & $\Delta_2 \!\geq\! 2J $ \\
&$ -\frac{1}{2}J\!\leq\!s\!\leq\! \frac{3}{2}J$ & $\Delta_1 \smpl  \Delta_2 \!\geq\! 3J$ \\ \hline
 $(1,\frac{1}{2},1)$ & $s \smeq  \Delta_1 \smpl  \Delta_2 \smmi \frac{5}{2}J$ & $\Delta_1 \!\leq\! 2J$ \\
 & & $\Delta_2 \!\leq\! 2J $ \\
 &$ \frac{1}{2}J\!\leq\!s\!\leq\! \frac{3}{2}J$ & $\Delta_1 \smpl  \Delta_2 \!\geq\! 3J$ \\ \hline
 $(1,\frac{3}{2},1)$ & $s \smeq  -\Delta_1 \smmi \Delta_2 \smpl  \frac{7}{2}J$ & $\Delta_1 \!\leq\! 2J$ \\
 & & $\Delta_2 \!\leq\! 2J $ \\
 &$ \frac{1}{2}J\!\leq\!s\!\leq\! \frac{7}{2}J$ & $\Delta_1 \smpl  \Delta_2 \!\leq\! 3J$ \\ \hline
\end{tabular}
\end{ruledtabular}
\caption{List of all possible ground state spin-transitions up to $S \smeq  \frac{3}{2}$ and their corresponding peak spacing. Only transitions with $ \Delta S\smeq\pm\frac{1}{2} $ are considered. 
Last column shows the conditions on the nearest level spacings so that the states involved in the transition are the ground states. Note that the even cases involve three consecutive level spacings.}
\label{Table}
\end{table}

\begin{figure}[t]
\includegraphics[width=8cm,clip]{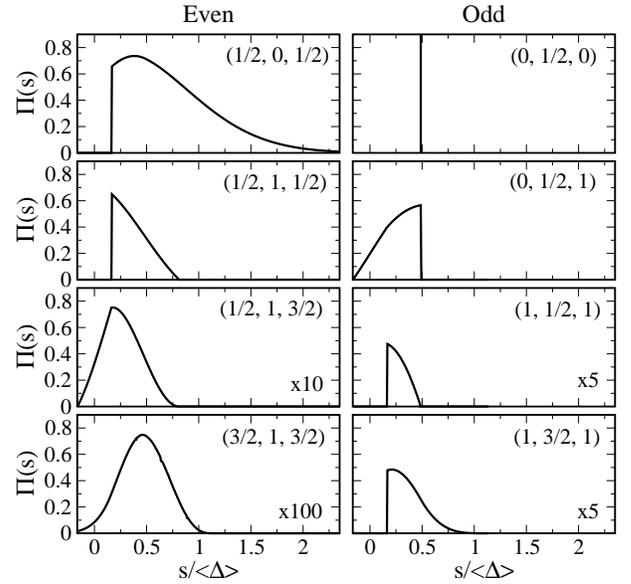} 
\caption{Partial contributions to the peak spacing distribution calculated using expressions in Appendix \ref{complicated} for the GOE and for $ J\smeq0.325\langle\Delta\rangle$. The distributions in the lower panels were multiplied by the indicated factor for the sake of comparison. Note the transitions involving $ S\smeq\frac{3}{2} $ add tails to both the even and the odd distribution and add a discontinuity to the latter at $ s\smeq\frac{1}{2}J $.}
\label{fig:partialdistribution}
\end{figure}

Unfortunately, this very simple approach is not enough to describe some important features of the peak spacing distribution for values of $ J/\langle\Delta\rangle\!\simeq\! 0.3$. States with $ S\smeq\frac{3}{2} $ have to be included.  Table \ref{Table} shows all possible spin transitions, the corresponding peak spacing, and the conditions on the level spacings for the states involved to be the ground states. We used the spin selection rule $ \Delta S\smeq\pm\frac{1}{2} $ upon the addition of an electron.  By a simple inspection of Table \ref{Table} we note several new features that appear in this analysis: (i) The even peak spacing distribution now depends not only on $ P_1( \Delta ) $ and $ P_2 ( \Delta_1, \Delta_2 ) $ but also on the joint probability distribution of three consecutive level spacings, $ P_3( \Delta_1, \Delta_2, \Delta_3 ) $---an accurate approximation to this quantity is derived in Section \ref{levelspacing}; (ii) even transitions with $ s\!<\!\frac{1}{2}J $ and odd transitions with $ s\!>\!\frac{3}{2}J $ are now possible; and (iii) a discontinuity appears in the odd distribution at $ s\smeq\frac{1}{2}J $.

An explicit expression for the peak spacing distribution is also possible in this case. We present it in Appendix \ref{complicated} as the calculation is rather lengthy. Here, we simply show the results for the GOE in Figure \ref{fig:partialdistribution}. Transitions $ (\frac{1}{2},1,\frac{3}{2}) $ and $ (\frac{3}{2},1,\frac{1}{2}) $  both give the same contribution, so only one is displayed; similarly for $ (0,\frac{1}{2},1)$ and $ (1,\frac{1}{2},0)$.



\begin{figure}[t]
\includegraphics[height=7.9cm,angle=0,clip]{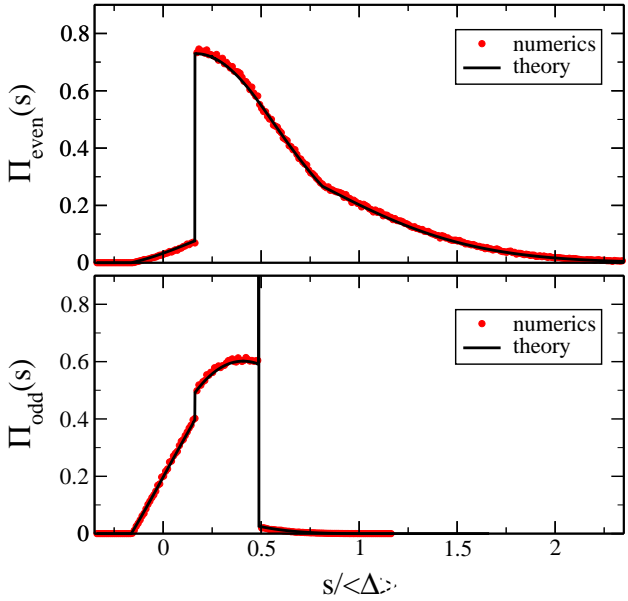}
\caption{(Color online) Comparison between the numerical result (dots) and the analytical result (solid line) for the peak spacing distribution in the Gaussian orthogonal ensemble with $ J\smeq0.325\langle\Delta\rangle $. The upper (lower) panel corresponds to the even (odd) case. The agreement is remarkably good.
}
\label{fig:orthogonalpeaks}
\end{figure}

\begin{figure}[t]
\includegraphics[height=7.9cm,clip]{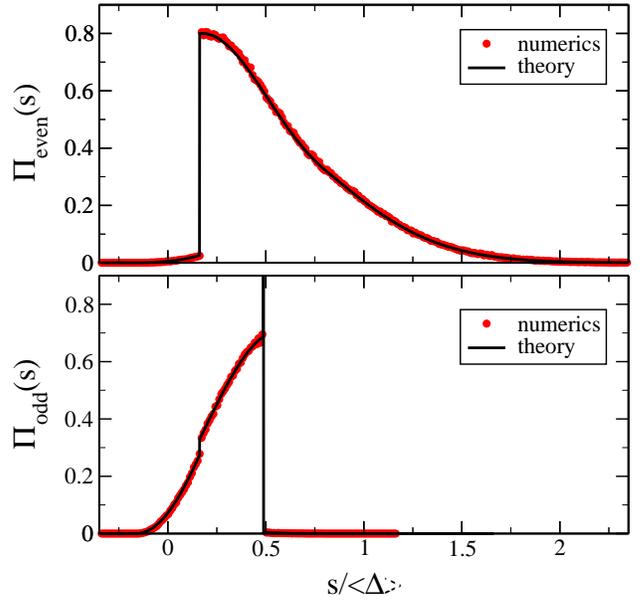}
\caption{(Color online) Same as in Fig.~\ref{fig:orthogonalpeaks} but for the Gaussian unitary ensemble.
The agreement here is even better due to the fact that the probability of large spin is lower in this case.}
\label{fig:unitarypeaks}
\end{figure}

A comparison of the peak spacing distribution, calculated in the approximation that allows no spin state higher than $ \frac{3}{2}$, to a numerical simulation \cite{ullmo01} is shown in Figures \ref{fig:orthogonalpeaks} and \ref{fig:unitarypeaks} for the GOE and GUE, respectively. The numerical calculation involved the diagonalization of $ 10^{6} $ random matrices of size $ 100\!\times\!100 $. Each random matrix corresponds to a different realization of the single particle levels in Eq.~(1), and the peak spacing distribution is inferred by histogramming the calculated peak spacings for these ``virtual quantum dots''. As can be seen from the figures, the agreement between the simulation and the analytic formula is excellent.

We have found that it is not necessary to consider higher
 spins for $ J/\langle \Delta \rangle $ as large as $0.4 $.
The analytic expression offers considerable advantage over the numerical calculation if, for example, it is necessary to compare the universal Hamiltonian model to an experimental distribution with the exchange constant as fitting parameter.

\section{Level Spacing Distributions: Simple Approximations
\label{levelspacing}}

In this Section we develop accurate approximations to the joint probability density of consecutive level spacings.  These expressions are needed to complete the expressions for the Coulomb blockade peak spacing distribution discussed above.

First it is helpful to recall Wigner's work \cite{mehta}. We are interested in $ P_1 ( \Delta ) $, the spacing between two consecutive levels near the center of the spectrum of a large $ M \!\times\! M $ random matrix. The exact result, derived by a lengthy and intricate calculation \cite{mehta}, involves an infinite product of eigenvalues of prolate spheroidal functions which are difficult to evaluate numerically. It is therefore of limited practical value.

Now, following Wigner, consider a $ 2 \!\times\! 2 $ real symmetric
Gaussian random matrix. Such a matrix has two eigenvalues, and
an elementary calculation shows that the spacing between them
is distributed according to 
\begin{equation}
P_{1{\rm WS}}^{ {\rm goe} } ( \Delta ) \smeq  
\frac{ \pi }{2} \Delta \exp \left( - \frac{ \pi }{ 4 } \Delta^2 \right)\;.
\label{eq:wsgoe}
\end{equation}
The scale is chosen so that the mean level spacing is unity,
\begin{equation}
\langle \Delta \rangle \smeq  
\int_0^{\infty}  \Delta\,
P_{1{\rm WS}}^{{\rm goe}} ( \Delta )\, d \Delta\smeq  1\,.
\label{eq:meanspacing}
\end{equation}
Eq.~(\ref{eq:wsgoe}) is the celebrated Wigner surmise. 
For the GUE, the surmise is similarly derived by considering a $ 2 \!\times\! 2 $ complex hermitian Gaussian random matrix, yielding
\begin{equation}
P_{1{\rm WS}}^{{\rm gue}} (\Delta) \smeq\frac{32}{ \pi^2 } \Delta^2 
\exp \left( - \frac{4}{\pi} \Delta^2 \right) \;.
\label{eq:wsgue}
\end{equation}
The Wigner surmise is known to be an excellent approximation to the level spacing distribution for the orthogonal ensemble where it has been checked against numerical calculations and rigorous bounds (see page 157 of Ref.\,\onlinecite{mehta}). For the unitary case, to our knowledge, the surmise has not been subject to comparably rigorous scrutiny, but it is generally believed to be very accurate \cite{guhr} (see also Fig.~8 below where the unitary Wigner surmise is compared to our numerical
calculation of the level spacing distribution).

Let us turn to the joint probability distribution of two consecutive level spacings, $ P_2 ( \Delta_1, \Delta_2 ) $. The exact result may be derived using the theory of Fredholm or Toeplitz determinants, as in Section \ref{consec} below. However, a more useful approximation may be derived by considering, in the spirit of the Wigner surmise, a $ 3 \!\times\! 3 $ real symmetric Gaussian random matrix (for the GOE). Using the standard joint probability distribution of the levels of a random matrix, we obtain
\begin{eqnarray}
P_{2; 3 \times 3}^{{\rm goe}} ( \Delta_1, \Delta_2 ) & \smeq  &
4 \sqrt{ \frac{2}{3 \pi} } a^\frac{5}{2} 
\Delta_1 \Delta_2 ( \Delta_1 \smpl  \Delta_2 ) \\
& \times &
\exp \left( - \frac{2a}{3} [ \Delta_1^2  \smpl  \Delta_2^2 \smpl  \Delta_1 \Delta_2 ]
\right) \nonumber
\label{eq:3x3goe}
\end{eqnarray}
with $ a \smeq  27/(8\pi)$.  The distribution has been normalized and the energy scale chosen so that $ \langle \Delta_1 \rangle \smeq  \langle \Delta_2 \rangle \smeq  1$. We conjecture that Eq.~(\ref{eq:3x3goe}) is an accurate approximation to the true joint spacing distribution for large GOE matrices.

A similar result for the GUE may be derived by considering $ 3 \!\times\! 3 $ complex hermitian Gaussian random matrices. We obtain
\begin{eqnarray}
P_{2; 3 \times 3}^{{\rm gue}} ( \Delta_1, \Delta_2 ) & \smeq  &
\frac{4\,b^4}{\pi\sqrt{3} }  \,
\Delta_1^2 \Delta_2^2 ( \Delta_1 \smpl  \Delta_2 )^2 
\label{eq:3x3gue}\\
&\times &
\exp \left( - \frac{2\,b}{3} 
[ \Delta_1^2 \smpl  \Delta_2^2 \smpl  \Delta_1 \Delta_2 ]
\right) \nonumber
\end{eqnarray}
with $ b \smeq  729/(128 \pi) $, which we conjecture is a good approximation to the true joint spacing distribution for the GUE.

There is no {\em a priori} justification for either Wigner's surmise or for our analogous conjectures. To test our conjectures we subjected them to many checks, some of which we describe here.  For definiteness we focus here on the unitary ensemble; but we have done similar tests in the orthogonal case.

\begin{figure}[t]
\includegraphics[height=6cm,clip]{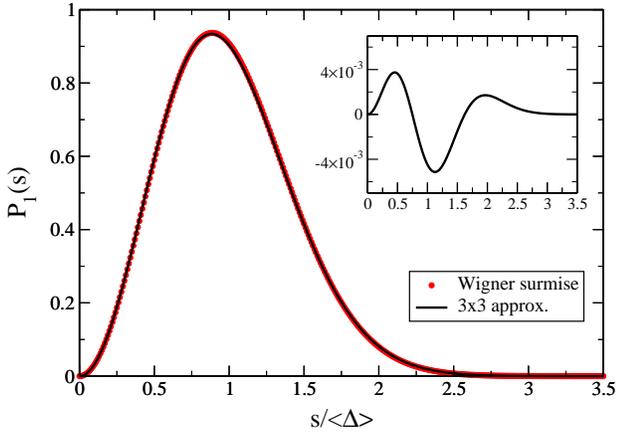}
\caption{(Color online). Comparison between the GUE Wigner surmise and the $ 3 \!\times\! 3 $
approximation Eq.~(\ref{eq:test1}). Inset: Difference between the latter and the former.The agreement is excellent.}
\label{fig:3by3wignersurmise}
\end{figure}

First, we note the relationship
\begin{equation}
P_1^{{\rm gue}} (\Delta_1) \smeq  
\int_{0}^{\infty}  P_2^{{\rm gue}} ( \Delta_1, \Delta_2 )\,d \Delta_2
\label{eq:integrali}
\end{equation}
between the exact single level spacing distribution $ P_1 $ and the exact
joint spacing distribution $ P_2 $. Performing the indicated integral
using our surmise Eq.~(\ref{eq:3x3gue}), we obtain 
\begin{eqnarray}
\label{eq:test1}
P_{1; 3 \times 3}^{{\rm gue}} ( \Delta_1 ) & \smeq  & \int_0^{\infty} 
P_{2; 3 \times 3}^{{\rm gue}} ( \Delta_1, \Delta_2 )\,d \Delta_2 \\
& \smeq  & \frac{9\, b^\frac{3}{2}}{2\sqrt{2}\pi} \Delta_1^2 
\exp\! \left( - \frac{b}{2} \Delta_1^2 \right)
h\! \left( \Delta_1 \sqrt{ \frac{b}{6} } \right) \nonumber
\end{eqnarray}
where 
\be
h (x) = - x^3 e^{- x^2} + \frac{3}{2} x e^{ - x^2 }
+  \sqrt{\pi}\Big( \frac{3}{4} \smmi x^2 \smpl x^4 \Big) 
\mathrm{Erfc}(x)
\label{eq:hdelta}
\ee
and $ \mathrm{Erfc}(x) $ is the complementary error function.

\begin{figure}[t]
\includegraphics[width=8cm,clip]{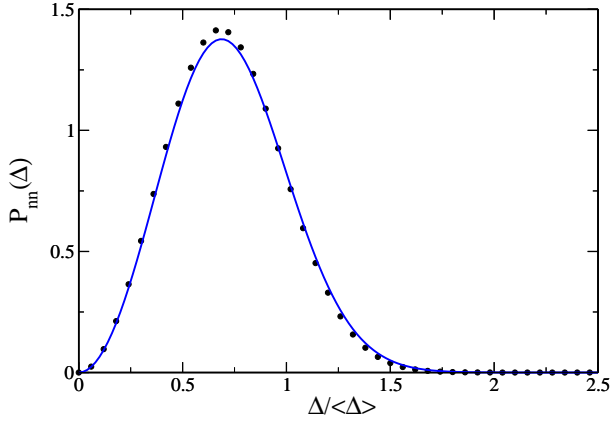} 
\caption{(Color online) The nearest neighbor spacing distribution $P_{{\rm nn}} (\Delta)$ calculated using the Toeplitz representation Eq.~(\ref{eq:pnntoeplitz}) with $n \!=\! 200$ (black points) is in good agreement with the generalized Wigner surmise approximation, Eq.~(\ref{eq:nnsurmise}) (blue curve).}
\label{fig:pnn}
\end{figure}

Fig.~\ref{fig:3by3wignersurmise} shows a comparison between the Wigner surmise and the $ 3 \!\times\! 3 $ approximation Eq.~(\ref{eq:test1}). The agreement is excellent over the entire region where the distribution has substantial weight and also for small spacings. In the tails, the relative error is larger but, of course, very small on an absolute scale. To be precise $99\%$ of the weight of $ P_{1{\rm WS}}^{{\rm gue}} ( \Delta ) $ lies in the range $ 0\! <\! \Delta\! <\! 2.11$. Over this range, $ P_{1{\rm WS}}^{{\rm gue}} $ and $ P_{1;3 \times 3}^{{\rm gue}} $ disagree by less than $3\%$.

%
%


Another quantity related to $ P_2 ( \Delta_1, \Delta_2 ) $ is the nearest neighbor spacing introduced by Forrester and Odlyzko in connection with studies of the zeros of the zeta function:\cite{ForrOdlyzko} the nearest neighbor is the closer of the level just above and the level just below a given level.
The distribution of the nearest neighbor spacing is, therefore,
\begin{eqnarray}
P_{{\rm nn}}^{{\rm gue}} ( \Delta ) & \smeq  & \!
\int_0^{\infty}\! d \Delta_1\! \int_0^{\Delta_1}\! \delta ( \Delta \smmi \Delta_2 )\,
P_2^{{\rm gue}} ( \Delta_1, \Delta_2 )\, d \Delta_2
\nonumber \\
& \smpl & 
\int_0^{\infty}\! d \Delta_1\!\int_{\Delta_1}^{\infty}\!\! \delta ( \Delta \smmi \Delta_1 )\,
P_2^{{\rm gue}} ( \Delta_1, \Delta_2 )\, d \Delta_2\,.
\nonumber \\
& &
\label{eq:nearestneighbour}
\end{eqnarray}
%
If we substitute the $3 \times 3$ approximation (\ref{eq:3x3gue})
into Eq.~(\ref{eq:nearestneighbour}), we obtain the $3 \times 3$ approximation
to the nearest neighbor distribution, 
$P^\mathrm{gue}_{\mathrm{nn};3\times3}(\Delta)\smeq b^{\frac{1}{2}} g_{\mathrm{nn}}\big((b/6)^{\frac{1}{2}}\Delta\big) $ where
\bea
g_{\mathrm{nn}}(x)&\smeq &\frac{27}{2\sqrt{2}\pi} x^2 e^{-12x^2}\big[18x + 84x^3
\nonumber\\
&&+ e^{9x^2} \pi^{\frac{1}{2}}(3\smmi4x^2\smpl4x^4)\mathrm{Erfc}(3x)\big]
\label{eq:nnsurmise}
\eea
Fig.~\ref{fig:pnn} shows a comparison between this $3 \times 3$ approximation to $P_{{\rm nn}}$ and the exact numerical computation in Section \ref{exact}. Again the agreement is very good except in the tails which have negligible weight. We have also confirmed that these two results are in good agreement with the exact numerical computation of $P_{{\rm nn}}$ using the method of Forrester and Odlyzko.\cite{ForrOdlyzko}

The covariance of consecutive spacings is also known \cite{mehta}.  The exact result for $ {\rm cov}(\Delta_1, \Delta_2)$ is $ -0.922 $ in the orthogonal ensemble and $-0.944  $ in the unitary ensemble; within our approximation for $P_2 $ they are found to be $ -8\pi/27\!\simeq\!-0.930$ and $-32(27\sqrt{3}\smmi8\pi)/729\!\simeq\!-0.950$, respectively. 

As a final check we can compare Eq.~(\ref{eq:3x3gue}) directly to the exact numerical computation of $P_2$ discussed in Section \ref{exact}. Again the agreement is reasonable (see Fig.~\ref{fig:ptwo} below).

Next, we turn to the joint probability distribution of
three consecutive spacings. To this end, we consider a $ 4 \!\times\! 4 $
real symmetric Gaussian random matrix. From the standard expression 
for the distribution of eigenvalues for a random matrix \cite{mehta}, we obtain
the normalized distribution
\begin{widetext}
\begin{equation}
{\cal P}_{3; 4 \times 4}^{{\rm goe}} ( \Delta_1, \Delta_2, \Delta_3 )
\! \smeq \!  \frac{8}{\sqrt{ \pi }} \Delta_1 \Delta_2 \Delta_3 ( \Delta_1 \smpl  \Delta_2 )
 ( \Delta_2 \smpl  \Delta_3 ) 
( \Delta_1 \smpl  \Delta_2 \smpl  \Delta_3 ) 
\exp [  - \frac{1}{4} \{ 2(\Delta_1+\Delta_2)^2+2
 (\Delta_2+ \Delta_3)^2 \smpl  (\Delta_1 \smpl  \Delta_3)^2\} ].
\nonumber 
\label{eq:unscaled4x4goe}
\end{equation}
\end{widetext}
Using this distribution, we find the average of the
middle spacing, $ f \!\equiv\! \langle \Delta_2 \rangle \!\approx\! 0.8388 $. 
A difficulty that now arises is that due to end effects in our
extremely finite sized matrix, $ f' \!\equiv\! \langle \Delta_1 \rangle \smeq  
\langle \Delta_3 \rangle \!\approx\! 0.9400$. We resolve this problem
by the standard procedure of ``unfolding'' \cite{BohigasLesHouches}, introducing the
rescaled distribution
\begin{equation}
P_{3; 4 \times 4}^{{\rm goe}} ( \Delta_1, \Delta_2, \Delta_3 ) \smeq  f'^2
f \,{\cal P}_{3; 4 \times 4}^{{\rm goe}} ( f' \Delta_1, f \Delta_2, 
f' \Delta_3 ).
\label{eq:4x4goe}
\end{equation}
We conjecture that $ P_{3; 4 \times 4}^{{\rm goe}} ( \Delta_1, \Delta_2, \Delta_3 ) $ is an accurate approximation to the true joint spacing distribution for the GOE.

\begin{figure}[t]
\includegraphics[height=6cm,clip]{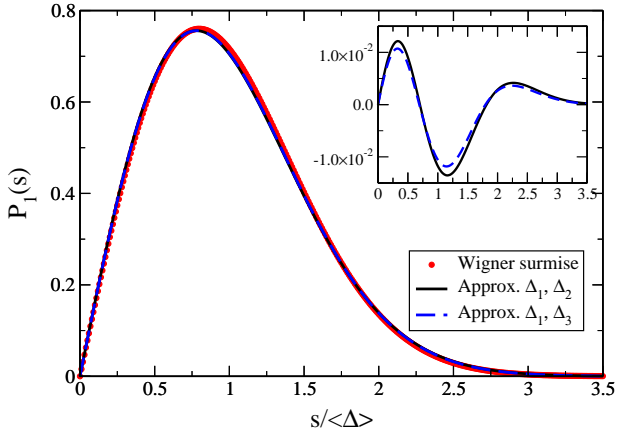}
\caption{(Color online). Comparison between the GOE Wigner surmise and the distribution obtained from the $ 4 \!\times\! 4 $ approximation Eq.~(\ref{eq:4x4goe}) by integrating out either $ \Delta_1 $ and $ \Delta_2 $ (solid line) or $ \Delta_1 $ and $ \Delta_3 $ (dashed line). Inset: Difference between each approximation and the Wigner surmise.}
\label{fig:4by4surmise}
\end{figure}

We have subjected this conjecture to tests similar to those applied to the distribution $ P_{2; 3 \times 3} $ above. For example, we can integrate out the end spacings $ \Delta_1 $ and $ \Delta_3 $ to obtain an approximation to the single level spacing distribution or we can integrate out the lower ones
 $ \Delta_1 $ and $ \Delta_2 $ to obtain a second approximation.
The two approximations are compared to each other and to the nearly exact Wigner surmise in Fig.~\ref{fig:4by4surmise} and are found to be accurate outside the tails.  On the basis of tests such as these we conclude that our conjectured distribution Eq.~(\ref{eq:4x4goe}) is sufficiently accurate for our purposes.

The corresponding distribution for the unitary ensemble can be similarly
derived by consideration of a $ 4 \!\times\! 4 $ complex hermitian matrix.
We obtain the normalized distribution
\begin{widetext}
\begin{equation}
{\cal P}_{3; 4 \times 4}^{{\rm gue}} ( \Delta_1, \Delta_2, \Delta_3 )  \smeq  
\kappa \Delta_1^2 \Delta_2^2 \Delta_3^2 ( \Delta_1 \smpl  \Delta_2 )^2  (
\Delta_2 \smpl  \Delta_3 )^2 (\Delta_1 \smpl 
\Delta_2 \smpl  \Delta_3 )^2 \exp [ - \frac{1}{4} \{ 2(\Delta_1+\Delta_2)^2+2
 (\Delta_2+ \Delta_3)^2 \smpl  (\Delta_1 \smpl  \Delta_3)^2\} ],
\label{eq:unscaled4x4gue}
\end{equation}
\end{widetext}
where $ \kappa \!\approx\!  0.4789$.
In this case we find $ h' \!\equiv\! \langle \Delta_1 \rangle \smeq  \langle \Delta_3 \rangle \!\approx\! 1.2288 $ whereas $ h \!\equiv\! \langle \Delta_2 \rangle \!\approx\! 1.1177$. Unfolding the distribution (\ref{eq:unscaled4x4gue}), we obtain
\begin{equation}
P_{3; 4 \times 4}^{{\rm gue}} ( \Delta_1, \Delta_2, \Delta_3 ) \smeq  h'^2
h \,{\cal P}_{3; 4 \times 4}^{{\rm gue}} ( h' \Delta_1, h\Delta_2, h' \Delta_3 )\,,
\label{eq:4x4gue}
\end{equation}
which we find to be an accurate approximation to the true joint spacing distribution for the GUE.

In summary, we have obtained simple expressions for the joint spacing distribution of two consecutive spacings, Eqs.~(\ref{eq:3x3goe}) and (\ref{eq:3x3gue}), and for three consecutive spacings, Eqs.~(\ref{eq:4x4goe}) and (\ref{eq:4x4gue}), for the orthogonal and unitary ensembles, respectively. Together with the Wigner surmise, Eqs.~(\ref{eq:wsgoe}) and (\ref{eq:wsgue}), these distributions are the random matrix theory tools needed to compute the Coulomb blockade peak spacing distribution in the paramagnetic limit. Using these tools, we obtained the peak spacing distributions already given in Section \ref{sec:peakspacing}.

\section{Level Spacing Distributions: Exact Results \label{exact}}

In this Section we derive a number of exact results in random matrix theory. These results allow us to certify the accuracy of the more useful simple approximations developed in the previous section. Some of the results, notably the new exact calculation of $P_{{\rm nn}}( \Delta )$ and the analysis of its asymptotic behavior, are also of intrinsic interest in random matrix theory.

The problem under consideration is the following: We are given a set of $n$ energy levels and $P(x_1, \ldots, x_n)$, their joint probability distribution. According to random matrix theory the energy level distribution is
\begin{equation}
P(x_1, \ldots, x_n) = {\cal N} 
\prod_{1 \leq i < j \leq n}
| x_i - x_j |^{\beta} 
\exp \left( - \alpha \sum_{i=1}^{n} x_i^2 \right),
\label{eq:rmt}
\end{equation}
where the exponent $\beta \smeq 1$, $2$, or $4$ corresponds respectively to the orthogonal, unitary, and symplectic ensembles of random matrix theory. ${\cal N}$ is a normalization constant, and $1/\sqrt{\alpha}$ is an energy scale set by the mean spacing between levels. 

>From the joint probability distribution $P(x_1, \ldots, x_n)$
we wish to calculate quantities such as the consecutive level
spacing distribution
\begin{eqnarray}
P_1 (t) & \equiv & {\cal K}_1 n (n-1)
\int_{{\rm out}} d x_3 \ldots 
\int_{{\rm out}} d x_n 
\nonumber \\
& &
P(x_1 \rightarrow - \frac{t}{2}, x_2 \rightarrow \frac{t}{2},
x_3, \ldots, x_n) \,,
\label{eq:levelspace}
\end{eqnarray}
where
\begin{equation}
\int_{{\rm out}} d x \equiv \int_{-\infty}^{-t/2} d x +
\int_{t/2}^{\infty} d x \;.
\label{eq:levelspaceout}
\end{equation}
This is the probability that one level is at $-t/2$, another at $t/2$, and none of the others are in between. We may place any of the $n$ levels at $-t/2$ and any of the remaining $(n-1)$ levels at $t/2$; this is the origin of the combinatoric prefactor in (\ref{eq:levelspace}). The normalization ${\cal K}_1$ ensures that
$\int_{0}^{\infty} d t P_1 (t) \smeq 1$.
Generally we wish to calculate the level spacing distribution in the limit $n \rightarrow \infty$. In addition to the level spacing, similar quantities of interest here are a more primitive quantity that we shall call the spacing determinant (defined below), the nearest neighbor spacing, and the joint probability distribution of two consecutive spacings.

The essential difficulty in random matrix calculations is the highly correlated nature of the probability distribution Eq.~(\ref{eq:rmt}). Level spacing calculations are particularly difficult due to the piecewise continuous nature of the integration domain. Nonetheless, exact formal expressions for these quantities have been derived. For example, the spacing determinant can be expressed as the Fredholm determinant of a particular integral operator $K$, the $n \rightarrow \infty$ limit of a certain $n \times n$ Toeplitz determinant, and the solution to an ordinary non-linear second-order differential equation. Explicit evaluation of the spacing determinant is then traditionally carried out numerically from the Fredholm determinant or the non-linear differential equation. In Section \ref{lsd} we find that direct evaluation of the Toeplitz determinant for large but finite $n$ provides a third efficient numerical method of calculating the spacing determinant.

In Sections \ref{joint}, \ref{nnspace}, and  \ref{consec} we derive new expressions for the consecutive level spacing $P_1 (\Delta)$, the nearest neighbor spacing $P_{{\rm nn}} (\Delta) $, and the joint distribution of two consecutive spacings $P_2 (\Delta_1, \Delta_2)$, expressing each of these quantities as a Toeplitz determinant.  These quantities may then be numerically evaluated by direct calculation of the determinants.  By contrast, in the conventional approach they are expressed in terms of the eigenvalues and eigenfunctions of the kernel $K$, which are harder to handle numerically.  The Toeplitz representation also makes it easy to obtain the large $\Delta$ asymptotic behavior: the previously unknown asymptotic behavior of $P_{{\rm nn}} (\Delta)$ is given in Section \ref{asymptotics}. For simplicity, we concentrate upon the unitary ensemble, $\beta \smeq 2$; the other ensembles will be studied in future work.

Toeplitz determinants arise in many contexts, among them signal processing \cite{numerical} and statistical mechanics \cite{fisher}.  Consequently, a great deal is known or conjectured about their asymptotic behavior and efficient numerical methods exist for their evaluation; that is the chief virtue of the Toeplitz expressions for $ P_1(\Delta) $, $P_{{\rm nn}}(\Delta)$, and $P_2(\Delta_1, \Delta_2)$ derived here.

\subsection{Level Spacing Determinant\label{lsd}}

The level spacing determinant is the probability that a band of width
$t$ is entirely void of energy levels,
\begin{equation}
E(t) \equiv 
\int_{{\rm out}} d x_1 \ldots
\int_{{\rm out}} d x_n
P(x_1, \ldots, x_n) \;,
\label{eq:eoft}
\end{equation}
where $ \int_{{\rm out}} d x $ is defined in Eq.~(\ref{eq:levelspaceout}).
Evidently $E(0) \smeq 1$, whereas $E(\infty) \smeq 0$.
By straightforward differentiation\cite{footnote1}
\begin{eqnarray}
\lefteqn{ F(t) \equiv - \frac{d E}{d t} }
\\
& & =  n 
\int_{{\rm out}} \!d x_2 \ldots
\int_{{\rm out}} \!d x_n\;
P(x_1 \rightarrow - \frac{t}{2}, x_2, \ldots, x_n) \;.
\nonumber 
\label{eq:foft}
\end{eqnarray}
$F$ is the probability that one level is at the edge of the
band while the others lies outside it. Evidently, 
$F(\infty) \smeq 0$ while $F(0) \smeq R_1(0)$ where
$R_1$ is the mean density of levels, or the one point
correlation, defined as
\begin{equation}
R_{1}(x) = n 
\int_{-\infty}^{\infty} \!d x_2 \ldots
\int_{-\infty}^{\infty} \!d x_n \,
P( x_1 \rightarrow x, x_2 \ldots x_n) \,.
\label{eq:rone}
\end{equation}

Differentiating once again, we find that the consecutive level spacing distribution, Eq.~(\ref{eq:levelspace}), is given by
\begin{equation}
P_1(t) = {\cal K}_1 \frac{d^2 E}{d t^2} \;.
\label{eq:poft}
\end{equation}
In deriving this equation we have assumed that the joint probability distribution $P(x_1, \ldots, x_n)$ vanishes if two levels coincide, and have also neglected a bulk term in which the derivative acts on the integrand in (\ref{eq:foft}).  The error is found to vanish as $n \rightarrow \infty$.  Eq.~(\ref{eq:poft}) shows that in principle $P_1$ is determined by $E$. In practice, $E$ is usually computed numerically, and hence Eq.~(\ref{eq:poft}) is not the best way to determine $P_1$. However, we may use this identity to show that the normalization constant is ${\cal K}_1 \smeq 1/R_1(0)$.

It is now useful to briefly recount how the exact formal expressions for $E(t)$ are obtained. It is more convenient for this purpose to use Dyson's circular unitary ensemble rather than the Gaussian unitary ensemble [Eq.~(\ref{eq:rmt})
with $ \beta \smeq 2$]. The circular unitary ensemble describes $n$ angles distributed around the unit circle according to the normalized distribution
\begin{equation}
P(\theta_1, \theta_2, \ldots, \theta_n) = 
\frac{1}{n!}
\frac{1}{(2 \pi)^n}
\prod_{1 \leq r < s \leq n}
| e^{i \theta_r} - e^{i \theta_s} |^2.
\label{eq:circensemble}
\end{equation}
It is well known that (up to an irrelevant scale factor that determines the mean level spacing) the local statistical properties of these angles are identical to those of energy levels governed by the GUE.

In the circular ensemble, $E(t)$ is the probability that all angles lie outside the arc $-t/2 < \theta < t/2$. To compute $E(t)$ it is helpful to rewrite
\begin{eqnarray}
\lefteqn{ P(\theta_1, \ldots, \theta_n) = 
\frac{1}{n!} 
\sum_{P,Q} (-1)^P (-1)^Q}
\\
& & 
\times \varphi_{P(1)} (\theta_1) \varphi_{Q(1)}^{*} (\theta_1) \ldots
\varphi_{P(n)} (\theta_n) \varphi_{Q(n)}^{*} (\theta_n) \,.
\nonumber
\label{eq:circdet}
\end{eqnarray}
Here $ \varphi_1 (\theta) \smeq 1/\sqrt{2 \pi}$, 
$\varphi_2 (\theta) \smeq e^{i \theta}/\sqrt{2 \pi}, \ldots, 
\varphi_n (\theta) \smeq e^{i (n-1) \theta}/\sqrt{2 \pi}$.
$P$ and $Q$ represent permutations of the integers $\{1, \ldots, n\}$ and $(-1)^{P,Q}$ is the parity of the permutation. This representation follows from (\ref{eq:circensemble}) by means of the Vandermonde identity
\begin{equation}
\prod_{1 \leq i < j \leq n}
( x_i - x_j ) = 
{\rm det} \hspace{2mm} M.
\label{eq:vandermonde}
\end{equation}
where $M$ is the $n \times n$ matrix with elements
$M_{ij} \smeq (x_j)^{i-1}$. 

If we define 
\begin{equation}
g_{rs} \equiv 
\int_{t/2}^{2 \pi - t/2} \!\!d \theta
\varphi_r (\theta) \varphi_s^{*} (\theta) =
\int_{t/2}^{2 \pi - t/2} \frac{d \theta}{2 \pi}
\exp[ i (r - s) \theta ],
\label{eq:ge}
\end{equation}
it follows
\begin{eqnarray}
E(t) &  = & \frac{1}{n!} \sum_{PQ} (-1)^P (-1)^Q 
g_{P(1), Q(1)} \ldots g_{P(n), Q(n)}
\nonumber \\
& = & 
\frac{1}{n!} \sum_{PR} (-1)^R
g_{P(1), R[P(1)]} \ldots g_{P(n), R[P(n)]}
\nonumber \\
& = & 
\frac{1}{n!} \sum_{PR} (-1)^R
g_{1, R(1)} \ldots g_{n, R(n)}.
\label{eq:esimplify}
\end{eqnarray}
To obtain the second line, we have introduced $Q \smeq RP$ and used $ (-1)^Q \smeq (-1)^P (-1)^R$.  The third line follows from a simple rearrangement of the terms in the summand. The $P$ sum is now trivial, and we find 
\begin{equation} 
E(t) = \det g \label{eq:etoeplitz} 
\end{equation} 
where $g$ is an $n \times n$ matrix with matrix elements given by (\ref{eq:ge}).

The mean spacing between angles is $2 \pi/n$ (show by computing the one point correlation $R_1$). Hence we introduce $\Delta$,
\begin{equation}
\Delta \equiv \frac{n}{2\pi} t \;,
\label{eq:rescale}
\end{equation}
as a measure of the band-width in units of the mean level spacing.
Making this change of variables, we may finally write
\begin{equation}
E(\Delta) = \lim_{n \rightarrow \infty} \det g 
\label{eq:limninfty}
\end{equation}
with $g$ given by Eqs.~(\ref{eq:ge}) and (\ref{eq:rescale}). 

An $n \times n$ Toeplitz matrix has elements
\begin{equation}
T_{rs} = \int_{0}^{2 \pi} 
\frac{d \theta}{2 \pi} f(\theta) \exp[ i (r-s) \theta]
\label{eq:toeplitz}
\end{equation}
that are controlled by a single function $f(\theta)$.  Comparing Eqs.~(\ref{eq:ge}) and (\ref{eq:toeplitz}), we conclude that the matrix $g$ is of the Toeplitz form with $f(\theta) \smeq 1 $ for $\pi \Delta/n \!<\! \theta \!<\! 2 \pi \smmi \pi \Delta/n$ and zero otherwise.  Hence Eq.~(\ref{eq:limninfty}) is the representation of the level spacing determinant in terms of a Toeplitz determinant.

To motivate the Fredholm representation, consider the 
eigenvectors of $g$,
\begin{equation}
\sum_{s=1}^{n} g_{rs} \psi_s^{(\alpha)} = 
\lambda^{(\alpha)} \psi_r^{(\alpha)} \;.
\label{eq:gevalue}
\end{equation}
Here $ \alpha \smeq 1, \ldots, n$ labels distinct eigenvectors
$\psi^{(\alpha)}$ and their eigenvalues $\lambda^{(\alpha)}$.
In the $n \!\rightarrow\! \infty$ limit, we set $s/n \!\rightarrow\! y$,
$r/n \!\rightarrow\! x$, and $(1/n) \sum_{s=1}^{n} \!\rightarrow\!
\int_0^{1} d y$. Explicitly evaluating $g_{rs}$ in (\ref{eq:ge}) and taking the $n \!\rightarrow\! \infty$ limit, we obtain the integral eigenvalue equation
\begin{equation}
\int_0^{1} \!dy \, K(x,y) \, \psi^{(\alpha)} (y) = 
\lambda^{(\alpha)} \psi^{(\alpha)} (x)
\label{eq:integral}
\end{equation}
with kernel
\begin{equation}
K(x,y) = \delta (x-y) 
- \frac{1}{\pi (x-y)} \sin \left[ \pi \Delta (x - y) \right].
\label{eq:kernel}
\end{equation}
Since the determinant of a matrix is the product 
of its eigenvalues, we may write
\begin{equation}
E( \Delta ) = \prod_{\alpha = 1}^{\infty} \lambda^{(\alpha)}
= \det K.
\label{eq:fredholm}
\end{equation}
This is the representation of $E$ as a Fredholm determinant, a virtue of which is that $n \!\rightarrow\! \infty$ has been explicitly taken.

Evaluation of the eigenvalues $\lambda^{(\alpha)}$ is facilitated by a remarkable connection between the integral kernel $K$ and the prolate spheroidal functions of classical mathematical physics \cite{mehta}.  It is found that the prolate spheroidal differential operator $L$ and the kernel $K$ commute and therefore have the same eigenfunctions. In practice it is easier to determine the eigenfunctions of $L$ and then to compute $\lambda^{(\alpha)}$ by applying $K$ to these eigenfunctions.

Finally, we  note that a third expression for $E(\Delta)$ was derived by Jimbo {\em et al.}, expressing it as the solution to a second-order non-linear differential equation \cite{jimbo}.

Each of these representations has proved useful in the past.  The asymptotic behavior of $E(\Delta)$ for large $\Delta$ [and hence of $P_1(\Delta)$ via Eq.~(\ref{eq:poft})] was derived by des Cloizeaux and Mehta by asymptotic analysis of the differential operator $L$ and independently by Dyson by means of an ingenious application of inverse scattering theory to the analysis of the kernel $K$ \cite{mehta}.  It can also be obtained from the Toeplitz representation using an asymptotic formula due to Widom \cite{widom,mehta}.

The first numerical computation of $E(\Delta)$ was based on evaluation of the Fredholm determinant, taking advantage of the connection to prolate spheroidal functions \cite{mehta}. Numerical solution of the non-linear differential equation was subsequently found to be more efficient 
\cite{jimbo}.

\begin{figure}[t]
\includegraphics[width=8cm,clip]{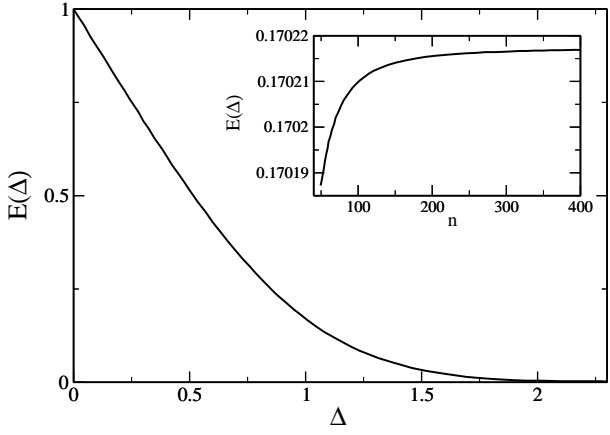} 
\caption{The spacing determinant $E(\Delta)$ calculated using
the Toeplitz representation Eq.~(\ref{eq:limninfty}) with $n\smeq200$. The inset
shows the dependence of $E(\Delta)$ on $n$ for $\Delta \smeq 1.$ Note the good precision obtained with modest numerical effort.}
\label{fig:det}
\end{figure}

Here we find that direct evaluation of the Toeplitz determinant Eq.~(\ref{eq:limninfty}) for large (but finite) $n$ also provides an accurate way to calculate $E(\Delta)$. Fig.~\ref{fig:det} shows a plot of $E(\Delta)$ calculated in this way with $n \smeq 200$; the inset shows the convergence of $E(\Delta)$ as a function of $n$ for $\Delta \smeq 1$.  The results for $n \smeq 100$ and $n \smeq 400$ differ only in the sixth significant figure. The convergence is slower for larger $\Delta$, but it is clear that for all values of $\Delta$ for which the distribution has significant weight, and to the accuracy needed for applications to quantum dots, Toeplitz determinants with $n$ equal to a few hundred should suffice.  Fig.~\ref{fig:det} is the main new result of this subsection.

\subsection{Consecutive Level Spacing\label{joint}}

In this sub-section we express the consecutive level spacing (sometimes simply called {\em the} level spacing) as a Toeplitz determinant. We then show that this expression can be used to compute $P_1(t)$ with good precision with modest numerical effort.

For simplicity let us suppose that there are $n \smpl 2$ angles on the
circle. $P_1(t)$ is the probability that one of these equals
$t/2$, another equals $-t/2$ and the others all lie outside the
range $-t/2 \!<\! \theta \!<\! t/2 $. Thus we must consider
\begin{eqnarray}
\lefteqn{P(\theta_1, \ldots, \theta_n, \theta_{n+1} \rightarrow
\frac{t}{2}, \theta_{n+2} \rightarrow - \frac{t}{2}) } \\
& & = \frac{1}{(n+2)!} \frac{1}{(2 \pi)^2} \frac{1}{(2 \pi)^n} 
| e^{it/2} - e^{-it/2} |^2 \times
\nonumber \\
& & \prod_{r=1}^{n}
| e^{i \theta_r} - e^{i t/2} |^2
| e^{i \theta_r} - e^{-it/2} |^2
\prod_{s=r+1}^n 
| e^{i \theta_r} - e^{i \theta_s} |^2 \,.
\nonumber 
\label{eq:ponecirc}
\end{eqnarray}
Comparing the two equivalent formulations of the circular unitary ensemble, Eqs.~(\ref{eq:circensemble}) and (\ref{eq:circdet}), we may write
\begin{eqnarray}
\lefteqn{ P(\theta_1, \ldots, \theta_n, \theta_{n+1} \rightarrow
\frac{t}{2}, \theta_{n+2} \rightarrow - \frac{t}{2}) } \label{eq:poneslater}
\\
& & = \frac{1 - \cos t}{2 \pi^2(n+2)!}
\sum_{P,Q} (-1)^P (-1)^Q \prod_{i=1}^n
\varphi_{P(i)} (\theta_i) \varphi_{Q(i)}^{*} (\theta_i) 
\nonumber \\
& & 
\quad \times \prod_{j=1}^n 4 
\Big[ 1 - \cos \big( \theta_j - \frac{t}{2} \big) \Big]
\Big[ 1 - \cos \big( \theta_j + \frac{t}{2} \big) \Big]
\;.
\nonumber
\end{eqnarray}
To compute $P_1(t)$ we must now integrate over $ \theta_1,
\ldots, \theta_n$ outside the range $ -t/2 < \theta < t/2$. 
Comparing Eq.~(\ref{eq:poneslater}) to (\ref{eq:circdet}), it is 
clear that if we define
\begin{eqnarray}
g_{rs}^{(1)} & \equiv & 
\int_{t/2}^{2 \pi - t/2} d \theta
\Big[ 1 - \cos \big( \theta - \frac{t}{2} \big) \Big]
\nonumber \\
& & \times
\Big[ 1 - \cos \big( \theta + \frac{t}{2} \big) \Big]
\varphi_r (\theta) \varphi_s^{*} (\theta)
\label{eq:gone}
\end{eqnarray}
then
\begin{equation}
P_1(t) = {\cal K}_1 
\frac{1}{2 \pi^2} (1 - \cos t)
\det g^{(1)} \;.
\label{eq:unscaledpone}
\end{equation}
The constant ${\cal K}_1$ is $1/R_1 \smeq (2\pi/n)$ [see the
discussion following Eq.~(\ref{eq:poft})]. Further, let us work with
$\Delta$, the spacing in units of the mean level spacing, 
rather than $t$; then
\begin{equation}
P_1 (\Delta) = 
\lim_{n \rightarrow \infty}
\frac{2}{n^2} \Big(1 - \cos \frac{2 \pi \Delta}{n} \Big)
\det g^{(1)} \;.
\label{eq:ponedet}
\end{equation}
Note that the distribution is normalized and the energy scaled so that the mean level-spacing is unity.
Eq.~(\ref{eq:ponedet}) is the main result of this subsection. It expresses the consecutive level spacing as a Toeplitz determinant. 

\begin{figure}[t]
\includegraphics[width=8cm,clip]{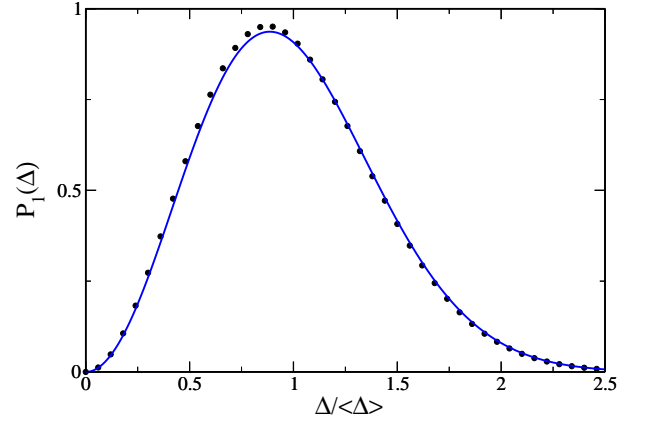} 
\caption{(Color online) The level spacing $P_1(\Delta)$ calculated
using the Toeplitz representation Eq.~(\ref{eq:ponedet}) with
$n \smeq 200$ (black points). Also plotted is the Wigner 
surmise (blue curve), known to give an excellent approximation
to the true distribution.}
\label{fig:pone}
\end{figure}

Fig.~\ref{fig:pone} shows a plot of $P_1(\Delta)$ computed by evaluation of the Toeplitz determinant for $n \smeq 200$ (black points). For comparison we have also plotted the Wigner surmise (blue curve) which is known to give an excellent approximation to the true spacing distribution. Fig.~\ref{fig:pone} shows that the $n \smeq 200$ approximation is adequate for any application to quantum dots.


\subsection{Nearest Neighbor Spacing\label{nnspace}}

A given level has two neighbors: one above, the other below.  Recently, Forrester and Odlyzko\cite{ForrOdlyzko} analyzed the distribution of distance to the nearer of the two neighbors, pointing out that this distribution is distinct from the consecutive level spacing distribution, the traditional object of study in random matrix theory. Forrester and Odlyzko adapted the differential equation method \cite{jimbo} in their study; here, we show that the distribution may also be expressed as a Toeplitz determinant.

As a preliminary, we introduce a more primitive quantity,
\begin{equation}
E_{{\rm nn}} \equiv
n \int_{{\rm out}} \!\!d x_2 \ldots 
\int_{{\rm out}} \!\!d x_n \,
P ( x_1 \rightarrow 0, x_2, \ldots, x_n),
\label{eq:enn}
\end{equation}
the probability that one level is at zero energy while the others
lie outside a band of width $2t$. Here
\begin{equation}
\int_{{\rm out}} d x = 
\int_{-\infty}^{-t} d x +
\int_{\infty}^{t} d x \;.
\label{eq:nnout}
\end{equation}
Evidently, $E_{{\rm nn}} ( \infty ) \smeq 0$ and
$E_{{\rm nn}} (0) \smeq R_1 (0)$, where $R_1$ is the one point correlation function defined in Eq.~(\ref{eq:rone}).

By straightforward differentiation we find 
\begin{eqnarray}
\lefteqn{ - \frac{d E_{{\rm nn}}}{d t} = 
n (n-1) \int_{{\rm out}} \!\!d x_3 \ldots \int_{{\rm out}} \!\!d x_n }
\nonumber \\
& &  \times \Big\{ 
P(0, t, x_3, \ldots, x_n ) 
+ P( 0, -t, x_3, \ldots, x_n ) \Big\}
\nonumber \\
& & \equiv R_1(0) P_{{\rm nn}} (t)
\label{eq:pnn}
\end{eqnarray}
where the normalization 
constant ensures that
\begin{equation}
\int_0^{\infty} d t \, P_{{\rm nn}} (t) = 1 \;.
\label{eq:nnnorm}
\end{equation}

As an application of these identities, consider the Poisson model: $n$ independent levels, each distributed uniformly
over the interval $ [-n/2, n/2] $. In this case it is trivial to 
calculate $E_{{\rm nn}}(t)$ and, upon differentiation, to find
\begin{equation}
\lim_{n \rightarrow \infty} \, P_{{\rm nn}} (t)_{\rm poisson} = 2 \exp (-2t) \;.
\label{eq:nnpoisson}
\end{equation}
By contrast, the consecutive level spacing is \cite{footnote2}
\begin{equation}
P_1 (t)_{\rm poisson} = \exp(-t) \;.
\label{eq:spacepoisson}
\end{equation}
Thus the Poisson model, which is believed to describe the 
spectral statistics of integrable models, provides an
explicit illustration of the distinction between consecutive
level spacing and nearest-neighbor level spacing distributions.

Now let us derive an expression for $E_{{\rm nn}} (t)$, Eq.~(\ref{eq:enn}),
within random matrix theory.
Once again, it is convenient to work with the circular ensemble,
and to suppose that of $(n+1)$ angles, one lies at zero, while the others
lie outside the arc $ - t < \theta < t$. To this end, we must consider
\begin{eqnarray}
\label{eq:pnncue}
\lefteqn{ (n+1) P(\theta_1, \ldots, \theta_n, \theta_{n+1} \rightarrow 0) }\\
& & = \frac{1}{n!} \frac{1}{2 \pi} \frac{1}{(2 \pi)^n}
\prod_{r = 1}^{n} | e^{i \theta_r} - 1 |^2
\prod_{1 \leq r < s \leq n} 
| e^{i \theta_r} - e^{i \theta_s} |^2.
\nonumber
\end{eqnarray}
Comparing the two equivalent forms of the circular unitary ensemble,
Eqs.~(\ref{eq:circensemble}) and (\ref{eq:circdet}), we may write
\begin{eqnarray}
\label{eq:pnnperm}
\lefteqn{(n+1) P(\theta_1, \ldots, \theta_n, 0) 
= \frac{1}{2 \pi} \frac{1}{n!} \sum_{P,Q} (-1)^P (-1)^Q} 
\\
& & \hspace*{2.0cm} \times 
\prod_{i=1}^n 2 \varphi_{P(i)} (\theta_i) \varphi_{Q(i)}^{*} (\theta_i) (1 - \cos \theta_i) \, .
\nonumber
\end{eqnarray}
Now, following the reasoning of the two preceding subsections leads to
the result
\begin{equation}
E_{{\rm nn}} ( \Delta) = \lim_{n \rightarrow \infty} 
\frac{1}{2 \pi} \det g_{{\rm enn}},
\label{eq:enntoeplitz}
\end{equation}
where $g_{{\rm enn}}$ is an $n \times n$ matrix of the standard
Toeplitz form, Eq.~(\ref{eq:toeplitz}), with the generating function $f(\theta)$ given by
\begin{equation}
f_{{\rm enn}} ( \theta ) = \left\{
\begin{array}{ll} 
2 (1 - \cos \theta),
& {\rm for} \hspace{2mm} \frac{2 \pi \Delta}{n} < \theta 
< 2 \pi - \frac{2 \pi \Delta}{n}
\\
0, & {\rm otherwise}
\end{array}
\right. \;.
\label{eq:ennfn}
\end{equation}

The analysis of $ P_{{\rm nn}}$ is entirely similar. The end result is
\begin{equation}
P_{{\rm nn}} ( \Delta) = \lim_{n \rightarrow \infty} 
\frac{4}{n^2} \left[ 1 - \cos \frac{2 \pi \Delta}{n} \right]
\det g_{{\rm pnn}}
\label{eq:pnntoeplitz}
\end{equation}
where $ g_{{\rm pnn}}$ is an $n \times n$ Toeplitz matrix generated by
\begin{equation}
f_{{\rm enn}} ( \theta ) = \left\{
\begin{array}{ll} 
4 (1 - \cos \theta) \left(1 - \cos \left[ \theta - \frac{2 \pi \Delta}{n}
\right] \right), \\
& \hspace*{-3.0cm} {\rm for} \hspace{2mm} \frac{2 \pi \Delta}{n} < \theta 
< 2 \pi - \frac{2 \pi \Delta}{n}
\\
0, & \hspace*{-3.0cm} {\rm otherwise}
\end{array}
\right. \;.
\label{eq:pnnfn}
\end{equation}

Eqs.~(\ref{eq:enntoeplitz})-(\ref{eq:pnnfn}) express $E_{{\rm nn}}(\Delta)$ and $P_{{\rm nn}} (\Delta)$ in terms of Toeplitz determinants and constitute the main result of this subsection. Fig.~\ref{fig:pnn} shows a plot of $P_{{\rm nn}} (\Delta)$ calculated by evaluation of Toeplitz determinants of size $ n \smeq 200 $ (black points). The result is in agreement with the earlier calculation by Forrester and Odlyzko\cite{ForrOdlyzko} based on the method of Jimbo {\em et al}\cite{jimbo}.

\subsection{A Pair of Consecutive Spacings\label{consec}}

The joint probability density for two consecutive spacings is defined
as 
\begin{eqnarray}
P_2( \tau, \xi ) & \equiv & {\cal K}_2 n (n-1)(n-2) 
\int_{{\rm out}} d x_4 \ldots \int_{{\rm out}} d x_n
\nonumber \\
& & 
\times P(x_1 \rightarrow 0, x_2 \rightarrow \tau, x_3 \rightarrow - \xi,
x_4, \ldots, x_n)
\nonumber \\
& &
\label{eq:ptwodefn}
\end{eqnarray}
with the excluded region given by
\begin{equation}
\int_{{\rm out}} d x \equiv
\int_{-\infty}^{-\xi} d x +
\int_{\tau}^{\infty} d x \;.
\label{eq:ptwoout}
\end{equation}
The normalization constant ${\cal K}_2$ is needed to ensure that
\begin{equation}
\int_{0}^{\infty} \!d \tau
\int_{0}^{\infty} \!d \xi
\,\, P_2 (\tau, \xi) = 1 \;.
\label{eq:ptwonorm}
\end{equation}

To determine the normalization it is convenient to 
introduce
\begin{equation}
E_2(\tau,\xi) = n 
\int_{{\rm out}} \!\!d x_2 \ldots
\int_{{\rm out}} \!\!d x_n 
\,P(x_1 \rightarrow 0, x_2, \ldots, x_n).
\label{eq:etwodefn}
\end{equation}
Evidently, $E_2(\infty, \xi) \smeq E_2(\tau, \infty) \smeq 0$
and $E_2 (0,0) \smeq R_1(0)$, the one point correlation.
It is also easy to see that
\begin{equation}
P_2 (\tau, \xi) = \frac{\partial^2}{\partial \tau 
\partial \xi} E_2 (\tau, \xi)
\label{eq:ptwoderiv}
\end{equation}
and hence ${\cal K}_2 \smeq 1/R_1(0)$. 

Again, in practice it is more convenient to work with the circular unitary ensemble and to compute the probability that of $(n+1)$ angles, one lies at $\theta \smeq 0$, another at $\tau$, a third at $-\xi$ and none of the others lie on the arc $- \xi < \theta < \tau$. It is also convenient to define $\tau \smeq 2 \pi \Delta_2/n$ and $\xi \smeq 2 \pi \Delta_1/n$; $\Delta_1$ and $\Delta_2$ are the level spacings in units where the mean spacing is one.

The calculations closely parallel those in the preceding 
subsections. The end result is that
\begin{eqnarray}
\label{eq:ptwotoeplitz}
\lefteqn{
P_2 (\Delta_1, \Delta_2 )  = \lim_{n \rightarrow \infty}
\frac{8}{n^3} 
\Big[ 1 - \cos \Big( \frac{2 \pi \Delta_1}{n} \Big) \Big] }
 \\[0.04in]
& &
\Big[ 1 - \cos \Big( \frac{2 \pi \Delta_2}{n} \Big) \Big]
\Big[ 1 - \cos \Big\{ \frac{2 \pi}{n} ( \Delta_1 +
\Delta_2 ) \Big\} \Big] \det g^{(2)}.
\nonumber 
\end{eqnarray}
Here $g^{(2)}$ is an $n\times n$ Toeplitz matrix generated by
\begin{equation}
f_{{\rm enn}} ( \theta ) = \left\{
\begin{array}{ll} 
8 \left[ 1 - \cos \theta \right]
\left[ 1 - \cos \left( \theta + \frac{2 \pi}{n} \Delta_1 \right) \right] 
\\ \hspace*{1.0cm} \times 
\left[ 1 - \cos \left( \theta - \frac{2 \pi}{n} \Delta_2 
\right) \right] 
, \\
& \hspace*{-3.5cm} {\rm for} \hspace{2mm} 
\frac{2 \pi}{n} \Delta_2 < \theta < 2 \pi - \frac{2 \pi}{n} \Delta_1
\\
0, & \hspace*{-3.5cm} {\rm otherwise}
\end{array}
\right. \;.
\label{eq:ptwofn}
\end{equation}

\begin{figure}[t]
\includegraphics[width=8cm,clip]{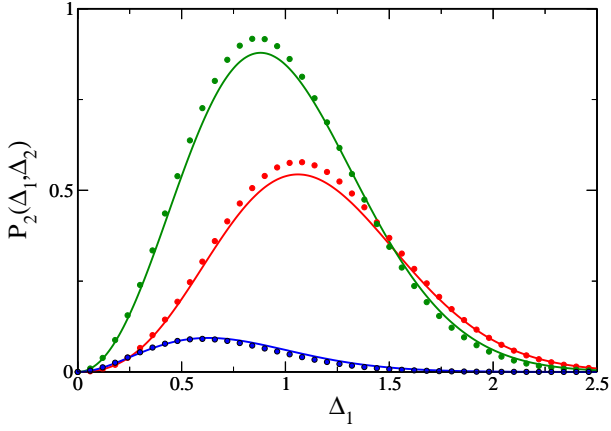} 
\caption{(Color online) The joint spacing distribution $P_2 (\Delta_1, \Delta_2)$ in the GUE,
calculated using the Toeplitz representation Eq.~(\ref{eq:ptwotoeplitz}) with $n \smeq 200$, as a function of $ \Delta_1 $ while $\Delta_2$ is held
fixed at 0.5 (red points), 1.0 (green points), and 2.0 (blue points).
The curves correspond to the generalized Wigner surmise approximation,
Eq.~(\ref{eq:3x3gue}).}
\label{fig:ptwo}
\end{figure}

Fig.~\ref{fig:ptwo} is a plot of $P_2(\Delta_1, \Delta_2)$ calculated using Eq.~(\ref{eq:ptwotoeplitz}) with $n \smeq 200$.  Fig.~\ref{fig:ptwo} and the Toeplitz representation of $P_2(\Delta_1, \Delta_2)$ in (\ref{eq:ptwotoeplitz}) and (\ref{eq:ptwofn}) are the main results of this subsection.  Mehta gives an expression for $P_2$ in terms of eigenvalues and eigenfunctions of the kernel, Eq.~(\ref{eq:kernel}), and tabulates $P_2$ for a small number of $(\Delta_1, \Delta_2)$ values.\cite{MehtaApp41} Our numerical scheme allows us to replicate and extend those results with only modest computational effort.

\subsection{Asymptotics\label{asymptotics}}

In this sub-section we employ the Toeplitz representation to obtain large $ \Delta $ asymptotic expressions for $E_{{\rm nn}}(\Delta) $ and $P_{{\rm nn}}(\Delta)$, analogous to the known results\cite{mehta} for $E(\Delta)$ and the consecutive level spacing $P_1(\Delta)$.  From a practical point of view these expressions are not very useful because the level spacing distributions have negligible weight in the tails. Nonetheless they are of some theoretical interest as they represent one limit in which it is possible to obtain exact analytic results for the level spacing problem.  Moreover, it is precisely when $ \Delta \gg 1$ that numerical computation of the distribution is most difficult.

The asymptotic results are obtained by straightforward application
of a formula due to Widom\cite{widom}. Consider an $n \times n$ Toeplitz
matrix $T$ with generating function $f(\theta)$ 
non-zero and positive on its support $ \alpha < \theta < 2 \pi - \alpha$,
and zero outside this arc. According to Widom, for large $n$ 
\begin{eqnarray}
\label{eq:widom}
\det T & \approx &
\left( n \sin \frac{\alpha}{2} \right)^{-1/4}
\left( \cos \frac{\alpha}{2} \right)^{n^2}
\\
& & \times
\exp( 2 b )
\exp( 2 F_0 ) 
\exp \left( \frac{1}{2} \sum_{k=1}^{\infty} k F_k F_{-k} \right) \;.
\nonumber 
\end{eqnarray}
Here
\begin{equation}
b = \frac{1}{24} \ln 2 + \frac{3}{2} \zeta'(-1)
\approx - 0.219250583,
\label{eq:bconst}
\end{equation}
and the function 
\begin{equation}
F(\theta) = f \left( 2 \cos^{-1} \left[ 
\cos \frac{\alpha}{2} \cos \theta \right] \right)
\label{eq:capf}
\end{equation}
is defined over the the range $ 0 \!< \! \theta  \!< \! 2 \pi$.
$F_0$ and $F_k$ are Fourier coefficients in the expansion
\begin{equation}
\ln F(\theta) = \sum_{k = - \infty}^{\infty} 
F_k \exp ( i k \theta ).
\label{eq:fourier}
\end{equation}

For our application $f(\theta) \smeq 2 (1 - \cos \theta)$ and
$\alpha \smeq 2 \pi \Delta/n$ with $1 \!\ll\! \Delta \ll n $ 
[Eqs.~(\ref{eq:enntoeplitz}) and (\ref{eq:ennfn})].
Thus 
\begin{equation}
F( \theta ) = 4 \left( 1 - \cos^2 \frac{\alpha}{2} \cos^2 \theta 
\right).
\label{eq:capfnn}
\end{equation}
The Fourier coefficients of $\ln F$ may be determined by contour
integration around the unit circle. We find
\begin{eqnarray}
F_0 & = & 4 \ln \left( \cos \frac{ \alpha}{2} \right) 
- 2 \ln \left(1 - \sin \frac{\alpha}{2} \right),
\nonumber \\
F_k & = & - \frac{2}{|k|} 
\left[ \frac{1 - \sin(\alpha/2) }{\cos (\alpha/2)} \right]^{|k|}.
\label{eq:fouriernn}
\end{eqnarray}
The expression for $F_k$ applies provided $k$ is non-zero 
and even. For odd $k$, $F_k \smeq 0$.

Substituting Eq.~(\ref{eq:fouriernn}) in (\ref{eq:widom}),
performing the $k$-sum, noting that $ \alpha \smeq 2 \pi \Delta/n$,
and taking the large $n$ limit, we finally obtain
\begin{eqnarray}
\lefteqn{ \frac{1}{n} \det g_{{\rm enn}} \approx } && \label{eq:asympdet}\\
&& \exp \left[ - 
\frac{\pi^2}{2} \Delta^2 +
2 \pi \Delta -
\frac{5}{4} \ln ( \pi \Delta ) + 
2 b - \ln 4 \right] \;.
\nonumber
\end{eqnarray}
$E_{{\rm nn}}$ follows directly from Eq.~(\ref{eq:enntoeplitz}), and  Eq.~(\ref{eq:pnn}) then implies that in the large $n$ limit
\begin{equation}
P_{{\rm nn}} ( \Delta ) \approx - \frac{d}{d \Delta} 
\left( \frac{1}{n} \det g_{{\rm enn}} \right)
\label{eq:rescaleenn}
\end{equation}
from which we obtain 
\begin{equation}
P_{{\rm nn}} \approx \frac{\pi}{4} 
\exp( 2 b )
\frac{1}{(\pi \Delta)^{1/4}}
\exp \left( - \frac{\pi^2 \Delta^2}{2} + 2 \pi \Delta \right).
\label{eq:asympnn}
\end{equation}

Eqs.~(\ref{eq:asympdet}) and (\ref{eq:asympnn}) are the asymptotic formulae that we sought. We see that $P_{{\rm nn}}(\Delta)$ has a much more rapid decay for large $\Delta$ than the consecutive level spacing; the latter is known to have the leading behavior $P_1(\Delta) \propto \exp( - \pi^2 \Delta^2/ 8 )$. This more rapid decay is already evident at quite modest values of $\Delta$ upon inspection of Figs. \ref{fig:pnn} and \ref{fig:pone}.

\section{Summary and Conclusion}

In this paper we have developed the relation between random matrix theory and electron-electron interaction in two dimensional electron gas quantum dots within the framework of the universal Hamiltonian model. We find the probability distribution of the spacing between conductance peaks in the Coulomb blockade regime for the standard orthogonal or unitary cases (the symplectic case is trivial, being essentially the same as the constant interaction model). For these cases there are departures from the constant interaction model even in the paramagnetic limit where $ J \!\ll\! \langle \Delta \rangle $. In this limit, the Coulomb blockade peak spacing distribution can be expressed in terms of the joint probability distribution of a few consecutive level spacings within RMT.  Making use of approximate formulae for the random matrix joint probability distributions, we obtain analytic formulae for the Coulomb blockade peak spacing distribution. These expressions reproduce earlier numerics \cite{ullmo01} and should facilitate future comparisons to experiments done at low temperature ($\kt\!<\!0.1\langle \Delta\rangle$) \cite{gonzalo,UsajB02}.

The simple approximate formulae for the joint probability distributions $P_2( \Delta_1, \Delta_2 )$ and $ P_3(\Delta_1, \Delta_2, \Delta_3)$ are obtained in Section \ref{levelspacing} by analogy to Wigner's famous surmise for the consecutive spacing distribution $P_1(\Delta)$.  In Section \ref{exact} we have also derived a number of new exact results in random matrix theory. These include a representation of the consecutive level spacing distribution $P_1(\Delta)$, the nearest neighbor spacing $P_{{\rm nn}}(\Delta)$, and the joint spacing distribution $P_2(\Delta_1, \Delta_2)$ as Toeplitz determinants. The Toeplitz representation enables efficient numerical computation of these distributions and allows us to infer the previously unknown large $\Delta$ asymptotics of $P_{{\rm nn}}(\Delta)$. Our primary motivation in deriving these exact results was to test the simple approximate formulae developed in Section \ref{levelspacing}, but the exact results may be of intrinsic interest from the viewpoint of fundamental random matrix theory also.

The derivation of the universal Hamiltonian model assumes that the wavefunctions are distributed according to a pure ensemble of random matrix theory. The problem of electron-electron interactions in the crossover between symmetry classes \cite{crossovers1,crossovers2,crossovers3,crossovers4,crossovers5,Murthy_crossovers,Murthy07a,Murthy07b} needs careful reconsideration.

\begin{acknowledgments}
We thank D. Ullmo for valuable conversations.
H. Mathur and D. Herman were supported by NSF Grant DMR 98-04983. H. U. Baranger and G. Usaj were supported in part by NSF Grant DMR-0506953. G.U. acknowledges support from CONICET (Argentina) and Fundaci\'on Antorchas (Grant 14169/21).
\end{acknowledgments}

\appendix

\section{Peak spacing distribution including $ S\smeq \frac{3}{2} $\label{complicated}}

The contribution to the probability distribution of the Coulomb blockade peak spacing of each spin transition is obtained by using the following standard procedure: if $ s\smeq f(x,y,z) $ is a function of random variables $ x $, $ y $, and $ z $, then its probability distribution is given by
\be
P(s)\smeq\int\!\!\int\!\!\int\! P(x,y,z) \,\delta\big(s\smmi f(x,y,z)\big)\,dx\,dy\,dz
\ee
where $ P(x,y,x) $ is the joint probability distribution of the three random variables.
Now, since in our case the function $ f(x,y,z) $ has different forms for different values of the variables (see Table \ref{Table}) we analyze each case separately and add the appropriate step functions ($ \Theta $).
The total spacing distribution is then the sum of each contribution. In what follows, both the even and the odd distribution are normalized to one.

\begin{widetext}
\subsection{Even distribution}
\be
P_{(\frac{1}{2},0,\frac{1}{2})}(s)\smeq\Theta\Big(s\smmi\frac{1}{2}J\Big)\int_0^\infty\!\int_0^\infty 
P_3\Big(x,s\smpl\frac{3}{2}J,z\Big)\Theta\Big(s\smpl x\smmi\frac{3}{2}J\Big)\Theta\Big(s\smpl z\smmi\frac{3}{2}J\Big)\,dx\,dz 
\ee
\be
P_{(\frac{1}{2},1,\frac{1}{2})}(s)\smeq\Theta\Big(s\smmi\frac{1}{2}J\Big)\Theta\Big(\frac{5}{2}J\smmi s\Big)\int_{s+\frac{1}{2}J}^\infty\!\int_{s+\frac{1}{2}J}^\infty 
P_3\Big(x,\frac{5}{2}J\smmi s,z\Big)\,dx\,dz 
\ee
\be
P_{(\frac{1}{2},1,\frac{3}{2})}(s)\smeq\Theta\Big(s\smpl\frac{1}{2}J\Big)\int_0^\infty\!dx
\int_0^{\mathrm{min}\left(2J,\frac{5}{2}J-s\right)}
P_3\Big(x,y,s\smpl\frac{1}{2}J\Big)\Theta\Big(x\smpl y\smmi3J\Big)\,dy
\label{eq:A4} 
\ee
\be
P_{(\frac{3}{2},1,\frac{3}{2})}(s)\smeq\int_0^{3J}\!dz\int_0^{\mathrm{min}\left(2J,3J-z\right)} 
P_3\Big(\frac{5}{2}J\smpl s\smmi y\smmi z,y,z\Big)\Theta\Big(z\smmi s\smpl\frac{1}{2}J\Big)\Theta\Big(s\smpl\frac{5}{2}J \smmi y\smmi z\Big)\,dy
\label{eq:A5}
\ee

\subsection{Odd distribution}
\be
P_{(0,\frac{1}{2},0)}(s)\smeq\delta\Big(s\smmi\frac{3}{2}J\Big)\int_{2J}^\infty\!\!\int_{2J}^\infty\! P_2(x,y)\,dx dy
\ee
\be
P_{(1,\frac{1}{2},0)}(s)\smeq\Theta\Big(s\smpl\frac{1}{2}J\Big)\Theta\Big(\frac{3}{2}J\smmi s\Big)
\!\int_{2J}^\infty\!  P_2\Big(s\smpl\frac{1}{2}J,y\Big)\Theta\Big(s\smpl y\smmi\frac{5}{2}J\Big)\,dy
\ee
\be
P_{(1,\frac{1}{2},1)}(s)\smeq\Theta\Big(s\smmi\frac{1}{2}J\Big)\!\int_0^{2J}\!\,P_2\Big(x,s\smpl\frac{5}{2}J\smmi x\Big)
\Theta\Big(x\smmi s\smmi\frac{3}{2}J\Big)\Theta\Big(\frac{5}{2}J\smpl s\smmi x\Big)\,dx
\ee
\be
P_{(1,\frac{3}{2},1)}(s)\smeq\Theta\Big(s\smmi\frac{1}{2}J\Big)\int_0^{2J}\!  P_2\Big(x,\frac{7}{2}J\smmi s\smmi x\Big)
\Theta\Big(\frac{7}{2}J\smmi s\smmi x\Big)\Theta\Big(x\smpl s\smmi\frac{3}{2}J\Big)\,dx
\ee

\subsection{Explicit expressions for the GOE case}
In the GOE case (see Fig.~\ref{fig:partialdistribution}), the integrals above can be done explicitly. For the even distribution we get
\bea
\nonumber
P_{(\frac{1}{2},0,\frac{1}{2})}^{\mathrm{goe}}(s)&\smeq& f\,\Theta\Big(s\smmi\frac{1}{2}J\Big)\left\{\Theta\Big(s\smmi \frac{3}{2}J\Big)4xe^{-3x^2}\left(\frac{2}{\pi^\frac{1}{2}}x\smmi e^{x^2}\left(2x^2\smmi1\right)\mathrm{Erfc}(x)\right)
 \smpl \Theta\Big(\frac{3}{2}J\smmi s\Big)4x e^{-3(2y^2+2yx+x^2)}\right.\\
\nonumber 
&&\left.\left(\frac{2}{\pi^\frac{1}{2}}(2y\smpl x)(1\smpl 2y(y\smpl x))\smpl
e^{(2y+x)^2}\left(1\smpl 2y^2(1\smpl y^2)\smpl2yx\left(1\smpl2y^2\right)\smmi2x^2\left(1\smpl y^2\right)\smmi4yx^3\right)\mathrm{Erfc}(2y\smpl x)\right)\right\}\\
\eea
where $x\smeq (s\smpl\frac{3}{2}J)f/\sqrt{3}$ and $y\smeq (\frac{3}{2}J\smmi s)f'/\sqrt{3}$, and
\bea
\nonumber
P_{(\frac{1}{2},1,\frac{1}{2})}^{\mathrm{goe}}(s)&\smeq&f\,\Theta\Big(s\smmi\frac{1}{2}J\Big)\Theta\Big(\frac{5}{2}J\smmi s\Big)4x e^{-3(2y^2+2yx+x^2)}
\Bigg(\frac{2}{\pi^\frac{1}{2}}(2y\smpl x)(1\smpl 2y(y\smpl x))\\
& & +
e^{(2y+x)^2}\left(1\smpl 2y^2(1\smpl y^2)\smpl2yx\left(1\smpl2y^2\right)\smmi2x^2\left(1\smpl y^2\right)\smmi4yx^3\right)\mathrm{Erfc}(2y\smpl x)\Bigg)
\eea
with $x\smeq(\frac{5}{2}J\smmi s)f/\sqrt{3}$ and $y\smeq (\frac{1}{2}J\smpl s)f'/\sqrt{3}$. 
The corresponding expressions for Eqs.~(\ref{eq:A4}) and (\ref{eq:A5}) are similar in form but too cumbersome to be presented here. 

The partial contributions to the odd distribution are given by 
\be
P_{(0,\frac{1}{2},0)}^{\mathrm{goe}}(s)\smeq\delta\Big(s\smmi\frac{3}{2} J\Big)e^{-j^2}\left(\frac{12^{\frac{1}{2}}}{\pi^{\frac{1}{2}}}je^{-3j^2}
\smmi \left(2j^2\smmi1\right)\mathrm{Erfc}\left(3^{\frac{1}{2}}j\right)\right)
\ee
with $ j\smeq J\sqrt{2b} $, and
\bea
\nonumber
P_{(1,\frac{1}{2},0)}^{\mathrm{goe}}(s)&\smeq&\sqrt{\frac{27b}{2\pi}}(j\smpl x)e^{-4(j\smpl x)^2}\left\{e^{-24j(5j\smmi x)}\left(2(11j\smmi x)\smmi e^{(x \smmi11j)^2}\pi^{\frac{1}{2}}\left(2(j\smpl x)^2\smmi1\right)\mathrm{Erfc}(11j\smmi x)\right)\Theta\Big(\frac{1}{2}J\smmi s\Big)\right. \\
\nonumber
&&\times\Theta\Big(\frac{1}{2}J\smpl s\Big)\left. +e^{-16j(5j\smpl x)}\left(2(9j\smpl x)\smmi e^{(x \smpl9j)^2}\pi^{\frac{1}{2}}\left(2(j\smpl x)^2\smmi1\right)\mathrm{Erfc}(9j\smpl x)\right)\Theta\Big(s\smmi \frac{1}{2}J\Big)\Theta\Big(\frac{3}{2}J\smmi s\Big)\right\}\\
\eea
\be
P_{(1,\frac{1}{2},1)}^{\mathrm{goe}}(s)\smeq\sqrt{\frac{54b}{\pi}}(5j\smpl x)e^{-4(21j^2\smpl6jx\smpl x^2)}\left(2(3j\smmi x)\smpl e^{(x \smmi3j)^2}\pi^{\frac{1}{2}}\left(2(5j\smpl x)^2\smmi1\right)\mathrm{Erf}(3j\smmi x)\right)\Theta\Big(s\smmi \frac{1}{2}J\Big)\Theta\Big(\frac{3}{2}J\smmi s\Big)
\ee
\bea
\nonumber
P_{(1,\frac{3}{2},1)}^{\mathrm{goe}}(s)&\smeq&\sqrt{\frac{54b}{\pi}}(7j\smmi x)e^{-4(49j^2\smmi10jx\smpl x^2)}\left\{e^{-48j^2}
\left(2(j\smpl x)\smpl e^{(x \smpl j)^2}\pi^{\frac{1}{2}}\left(2(x\smmi7j)^2\smmi1\right)\mathrm{Erf}(j\smpl x)\right)\Theta\Big(s\smmi \frac{1}{2}J\Big)\right. \\
\nonumber
&&\times\Theta\Big(\frac{3}{2}J\smmi s\Big)\left. +e^{16jx}\left(2(7j\smmi x)\smpl e^{(x \smmi7j)^2}\pi^{\frac{1}{2}}\left(2(x\smmi7j)^2\smmi1\right)\mathrm{Erf}(7j\smmi  x)\right)\Theta\Big(s\smmi\frac{3}{2}J\Big)\Theta\Big(\frac{7}{2}J\smmi s\Big)\right\}\\
\eea
with $ j\smeq J\sqrt{b/24} $ and $x\smeq s\sqrt{b/6} $.
\end{widetext}



\begin{thebibliography}{99}
\bibitem{review} L. P. Kouwenhoven, C. M. Marcus, P. L. McEuen, S. Tarucha,
R. M. Westervelt, and N. S. Wingreen, in {\em Mesoscopic Electron Transport},
edited by L.L. Sohn, L.P. Kouwenhoven, and G. Sch\"{o}n (Kluwer, New York,
1997), pp. 105-214.

\bibitem{IhnBook}
T. Ihn, \textit{Electronic Quantum Transport in Mesoscopic Semiconductor Structures} (Springer-Verlag, New York, 2004).

\bibitem{glazmanreview} I. L. Aleiner, P. W. Brouwer and L. I. Glazman,
Phys Rep {\bf 358}, 309 (2002).

\bibitem{oregreview} 
Y.~Oreg, P.~W.~Brouwer, X.~Waintal, and B.~I.~Halperin, in 
{\it Nano-Physics and Bio-Electronics: A New Odyssey}, 
edited by T.~Chakraborty, F.~Peeters, and U.~Sivan 
(Elsevier, Amsterdam, 2002) [cond-mat/01095413].

\bibitem{sivan} U. Sivan, R. Berkovits, Y. Aloni, O. Prus, A. Auerbach,
and G. Ben-Yoseph, Phys. Rev. Lett.{\bf 77}, 1123 (1996).

\bibitem{patel1} S. R. Patel, S. M. Cronenwett, D. R. Stewart, A. G. Huibers,
C. M. Marcus, C. I. Duru\"oz, J. S. Harris, K. Campman, and A. C. Gossard, 
Phys. Rev. Lett.{\bf 80}, 4522 (1998). 

\bibitem{patel2} S. R. Patel, D. R. Stewart, C. M. Marcus, M. G\"ok{\c c}eda{\v g},
Y. Alhassid, A. D. Stone, C. I. Duru\"oz, and J. S. Harris, Phys. Rev. Lett.
{\bf 81}, 5900 (1998).

\bibitem{simmel} F. Simmel, D. Abusch-Magder, D. A. Wharam, M. A. Kastner,
and J. P. Kotthaus, Phys. Rev. B {\bf 59}, R10441 (1999).

\bibitem{luscher01} S. L\"{u}scher, T. Heinzel, K. Ensslin, W. Wegscheider,
and M. Bichler, Phys. Rev. Lett.{\bf 86}, 2118 (2001).

\bibitem{ihn02}
S. Lindemann, T. Ihn, T. Heinzel, W. Zwerger, K. Ensslin, K. Maranowski, and A. C. Gossard, Phys. Rev. B 66, 195314 (2002).

\bibitem{fuhrer03}
A. Fuhrer, T. Ihn, K. Ensslin, W. Wegscheider, and M. Bichler, Phys. Rev. Lett. 91, 206802 (2003).

\bibitem{ong}
T. T. Ong, G. Usaj, H. U. Baranger, D. M. Higdon, S. R. Patel, and C. M. Marcus,
``Parity Effect in the Coulomb Blockade Peak Spacing Distribution of Quantum Dots,'' unpublished (2002).

\bibitem{BohigasLesHouches}
O. Bohigas, in \textit{Chaos and Quantum Physics}, edited by M.-J. Giannoni, A. Voros, and J. Zinn-Justin (North-Holland, Amsterdam, 1991).

\bibitem{EfetovBook}
K. Efetov, \textit{Supersymmetry in Disorder and Chaos} (Cambridge University Press, Cambridge, 1997).

\bibitem{mehta} 
M. L. Mehta, {\em Random Matrices} (Academic Press, San Diego, 1991).

\bibitem{kurland} I. L. Kurland, I. L. Aleiner and B. L. Altshuler,
Phys. Rev. B {\bf 62}, 14886 (2000).

\bibitem{brouwer}
P.~W.~Brouwer, Y.~Oreg, and B.~I.~Halperin, 
Phys. Rev. B {\bf 60}, R13977 (1999).

\bibitem{baranger}
H.~U.~Baranger, D.~Ullmo, and L.~I.~Glazman, 
Phys. Rev. B {\bf 61}, R2425 (2000).

\bibitem{AndKam98}
A. V. Andreev and A. Kamenev, Phys. Rev. Lett. \textbf{81}, 3199 (1998).

\bibitem{ullmo01} 
D. Ullmo and H. U. Baranger, Phys. Rev. B {\bf 64}, 245324 (2001).

\bibitem{gonzalo} G. Usaj and H.U. Baranger, Phys. Rev. B {\bf 64}, 201319(R) (2001).

\bibitem{UsajB02} G. Usaj and H. U. Baranger, Phys. Rev. B \textbf{66}, 155333 (2002).

\bibitem{AlhassidM02} Y. Alhassid and S. Malhotra, Phys. Rev. B \textbf{66}, 245313 (2002).

\bibitem{Orsay99}
See, however, O. Bohigas, P. Leboeuf, and M. J. S\'anchez,
Physica D \textbf{131}, 186 (1999), for the relation between some related random matrix quantities and fluctuations of the total energy of a Fermi gas.

\bibitem{ForrOdlyzko} 
P. J. Forrester and A. M. Odlyzko, Phys. Rev. E {\bf 54}, R4493 (1996).

\bibitem{guhr}
T. Guhr, A. Mueller-Groeling and H. A. Weidenmuller,
Phys. Reports {\bf 299}, 189 (1998).

\bibitem{numerical} W. H. Press, S. A. Teukolsky, W. T. Vettering and
B. P. Flannery, {\em Numerical Recipes}, 2$^{{\rm nd}}$ edition 
(Cambridge Univ Press, Cambridge, 1992).

\bibitem{fisher} M. E. Fisher and R. E. Hartwig, Adv. Chem. Phys. {\bf 15},
333 (1968); B. M. McCoy and T. T. Wu, {\em The Two-Dimensional Ising Model}
(Harvard Univ Press, Cambridge, Massachusetts, 1973).

\bibitem{footnote1}
A minor subtlety in taking the derivative is that $t$ appears only in the limits of the multidimensional integral Eq.~(\ref{eq:eoft}).

\bibitem{jimbo} M. Jimbo, T. Miwa, Y. Mori, and M. Sato, 
Jpn Acad Ser A, Math Sci {\bf 55}, 317 (1979); Physica D {\bf 1}, 80
(1980).

\bibitem{MehtaApp41}
See Appendix 41 in Ref.\,\protect\onlinecite{mehta}.

\bibitem{widom} B. Widom, Indiana Univ Math J {\bf 21}, 277 (1971).

\bibitem{footnote2}
The reader who wishes to verify this should use Eq.~(\ref{eq:levelspace}), not Eqs.~(\ref{eq:eoft}) or (\ref{eq:poft}). The Poisson distribution does not satisfy the condition, implicit in Eq.~(\ref{eq:poft}), that the probability of a degeneracy vanishes.

\bibitem{crossovers1}
H.-J. Sommers and S. Iida, Phys. Rev. E \textbf{49}, R2513 (1994).

\bibitem{crossovers2}
V. I. Fal'ko and K. B. Efetov, 
Phys. Rev. B \textbf{50}, 11 267 (1994); Phys. Rev. Lett. \textbf{77}, 912 (1996).

\bibitem{crossovers3}
S. A. van Langen, P. W. Brouwer, and C. W. J. Beenakker, 
Phys. Rev. E \textbf{55}, R1 (1997).

\bibitem{crossovers4}
S. Adam, P. W. Brouwer, J. P. Sethna, and X. Waintal, 
Phys. Rev. B \textbf{66}, 165310 (2002).

\bibitem{crossovers5}
D. A. Gorokhov and P. W. Brouwer, 
Phys. Rev. Lett. \textbf{91}, 186602 (2003); Phys. Rev. B \textbf{69}, 155417 (2004).

\bibitem{Murthy_crossovers}
G. Murthy, Phys. Rev. B \textbf{70}, 153304 (2004).

\bibitem{Murthy07a}
O. Zelyak, G. Murthy, and I. Rozhkov, preprint, arXiv:0704.0919 (2007).

\bibitem{Murthy07b}
G. Murthy, preprint, arXiv:0706.3406 (2007).






\end{thebibliography}
\end{document}